\documentclass[final,3p,times,twocolumn]{elsarticle}
\usepackage{graphicx}
\usepackage{amssymb}
\usepackage[utf8]{inputenc}
\usepackage{textcomp}

\usepackage{booktabs}
\usepackage{rotating}
\usepackage{multirow}
\usepackage{caption}
\usepackage{subcaption}

\usepackage{url,hyperref}

\usepackage{color}
\usepackage[usenames,dvipsnames,svgnames,table]{xcolor}

\usepackage{amsmath}

\journal{Journal of Nuclear Materials }

\begin{document}

\begin{frontmatter}



\title{The nanostructure evolution in Fe-C systems under irradiation at 560 K}


\author[sck,helsinki]{V.~Jansson\corref{cor1}}
\ead{ville.b.c.jansson@gmail.com}
\author[sck]{M. Chiapetto}
\author[sck]{L.~Malerba}

\cortext[cor1]{Corresponding author. Tel. +32 1433 3096, fax: +32 1432 1216.}

\address[sck]{Institute of Nuclear Materials Science, SCK$\bullet$CEN, Boeretang
200, 2400
{\sc Mol, Belgium}}
\address[helsinki]{Department of Physics, P.O. Box 43 (Pehr
Kalms gata 2), FI-00014 {\sc University of Helsinki, Finland}}

\begin{abstract}
This work extends our Object Kinetic Monte Carlo model for neutron irradiation-induced nanostructure evolution in Fe-C alloys to consider higher irradiation temperatures. The previous study concentrated on irradiation temperatures $<370$ K. Here we study the evolution of vacancy and self-interstitial atom (SIA) cluster populations at the operational temperature of light water reactors, by simulating specific reference irradiation experiments. 

Following our previous study, the effect of carbon on radiation defect evolution can be described in terms of formation of immobile complexes with vacancies, that in turn act as traps for SIA clusters. This dynamics is simulated using generic traps for SIA and vacancy clusters. The traps have a binding energy that depends on the size and type of the clusters and is also chosen on the basis of previously performed atomistic studies. 
The model had to be adapted to account for the existence of two kinds of SIA clusters, $\langle111\rangle$ and $\langle100\rangle$, as observed in electron microscopy examinations of Fe alloys neutron irradiated at the temperatures of technological interest. 

The model, which is fully based on physical considerations and only uses a few parameters for calibration, is found to be capable of reproducing the experimental trends, thereby providing insight into the physical mechanisms of importance to determine the type of nanostructural evolution undergone by the material during irradiation.

\end{abstract}


\end{frontmatter}

\frenchspacing

% \linenumbers

\section{Introduction}\label{sec:introduction}
The pressurized vessel of nuclear power plant reactors is generally an irreplaceable component, the integrity of which determines the lifetime of the installation. Prolonged exposure to irradiation causes hardening and embrittlement of the steel used to fabricate the vessel, which thereby might lose its capability to maintain integrity in case of an accident. It is therefore important to understand the origin of this hardening and embrittlement. It is known that changes in mechanical properties induced by irradiation can be understood in terms of changes in the micro- or nanostructure of the material. 

In this context, iron-carbon (Fe-C) alloys are archetypal model materials for low-alloy ferritic steels, such as those used for reactor pressure vessels of existing light water nuclear reactors. In particular, the generally-agreed mechanistic framework within which RPV steel hardening and embrittlement are understood involves the contribution of three damage components: (i) matrix damage, (ii) precipitates, and (iii) grain boundary de-cohesion due to segregation \cite{odette1998recent}. It is generally assumed that the first component, dominated by point-defects created in the matrix (iron), can be studied by reference to the behaviour of Fe-C \cite{buswell1995irradiation}. 

In this framework, this paper is a continuation of the work to develop a computer model for the nanostructure evolution in iron-carbon (Fe-C) alloys under irradiation, using the Object Kinetic Monte Carlo (OKMC) simulation technique. In our model, we study how the populations of SIA and vacancy clusters evolve over time and increasing damage accumulation in terms of displacements per atom (dpa). The clusters are described in the model by a set of parameters that define their stability, their diffusion properties, and their possibility of interacting with each other and with other defects and the pre-existing microstructure. 

In \cite{jansson2013simulation,jansson2014okmc} we presented a set of mechanisms and parameters valid for irradiated Fe-C systems at low temperature, below 370 K. At this temperature, transmission electron microscope examinations reveal the formation under neutron irradiation of only prismatic dislocation loops with $1/2\langle111\rangle$ Burgers vector \cite{malerba2011review,zinkle2006microstructure}. The same low temperature Fe-C system has also recently been studied using cluster dynamics modelling \cite{abe2013effect}. However, at higher temperature a vast majority of loops with $\langle100\rangle$ loops is observed \cite{malerba2011review,hernandez2010transmission}. The stability and the diffusion properties of $\langle100\rangle$ loops are significantly different from those of $1/2\langle111\rangle$ loops \cite{dudarev2008effect,arakawa2007observation}. The model must therefore be adapted to take this into account, in order to be applied to higher temperature conditions. We extend here the model to 
higher temperatures by taking as reference the irradiation experiment from the REVE campaign \cite{malerba2011review,hernandez2010transmission,lambrecht2008influence,lambrecht2009phd,lambrecht2009positron,bergner2010comparative}, where the irradiation temperature was 563--568 K. The ``pure'' Fe material in that experiment is estimated to contain less than 20 wt. ppm C. The average grain size is 250 $\mu$m and the dislocation density is reported to be $(7\pm2)\cdot10^{13}$ m$^{-2}$ The irradiation was performed in the in-pile section 2 of the CALLISTO loop of the BR2 reactor in SCK$\bullet$CEN in Mol, Belgium. The material was irradiated for 16 days until 0.19 dpa had accumulated. The fluence was $\sim$13.1$\cdot10^{19}$ n cm$^{-2}$, $E>1$ MeV \cite{malerba2011review}. This corresponds to a dpa rate of $1.37\cdot10^{-7}$ dpa/s.
After irradiation, the material was studied using positron annihilation spectroscopy (PAS) \cite{lambrecht2008influence,lambrecht2009phd,lambrecht2009positron,meslin2010characterization}, small-angle neutron scattering (SANS) \cite{bergner2010comparative} and transmission electron microscopy (TEM) \cite{hernandez2010transmission}. These different experimental techniques gives a thorough description of the nanostructure of the material in terms of defect densities and average vacancy and SIA cluster sizes.

% Horton?
% Another experiment was performed by L.L. Horton \textit{et al.}  \cite{horton1982tem}, where they irradiated Fe samples at differentiate temperature in a range of 455--1013 K. The samples accumulated between 0.51 dpa to 0.98 dpa. The fluence were between $2.25\cdot10^{25}$ to $4.26\cdot10^{25}$ neutrons/m$^{2}$. The fast fluence ($E>0.1$ MeV), from which the dpa vales were calculated, ranged from $0.69\cdot10^{25}$ to $1.32\cdot10^{25}$ neutrons/m$^2$. The dislocation density of the material was measured to be $<10^{11}$ m$^{-2}$ and the C content to 30 wt. ppm. After irradiation, the samples were studied with TEM and the general trend in their results are that the number of SIA clusters decrease with higher irradiation temperature, whereas the average SIA cluster size becomes larger. The large majority of the observed SIA clusters were of $\langle100\rangle$ type.

% The purpose of this study is to present a model for the nanostructure evolution in Fe-C under irradiation at temperatures above 370 K and thereby extend the already existing model for lower temperatures. We will verify the model by simulating the REVE irradiation experiment, which is the most complete study for high irradiation temperatures.

% The structure of this paper. Perhaps not necessary for such a short paper.

\section{Computation method}\label{sec:methods}

We use the LAKIMOCA code for our OKMC simulations \cite{domain2004simulation}. The approach we use is explained in detail  in \cite{jansson2013simulation} and \cite{jansson2014okmc}. 
For convenience, we highlight here the fundamental ideas on the method.

OKMC is a stochastic simulation technique that considers the migration and interactions of objects in a pre-defined system with parameterized probabilities. The objects represent in our case vacancy and SIA clusters. Their shapes are usually spherical, except for large SIA clusters, that are represented by toroids. Reactions, such as clustering or annihilation, take place when the objects overlap geometrically. The probability for the objects to perform a migration jump are given in terms of Arrhenius frequencies for thermally activated events, 
\begin{equation}
\Gamma_i = \nu_i\exp \left(\frac{-A_i}{k_B T}\right),
\end{equation}
where $\nu_i$ is the attempt frequency of event $i$, $A_i$ the activation energy, $k_B$ Boltzmann's constant and $T$ the temperature. Events are randomly chosen according to their probability, according to the Monte Carlo algorithm \cite{metropolis1953equation}. The simulated time is increased according to the residence time algorithm \cite{young1966monte} with
\begin{equation}\label{eq:residence}
\Delta t = \frac{1}{\sum_{i=1}^{N_{int}} \Gamma_i + \sum_{j=1}^{N_{ext}}P_j},
\end{equation}
where $N_{int}$ is the number of internal events such as defect jumps and $N_{ext}$ the number of external events, such as cascades or Frenckel pair creation, with $P_j$ being the probabilities for the external events. In the long term, Eq. \eqref{eq:residence} is indeed equivalent to 
\begin{equation}
\Delta t' = -\ln{R}\Delta t,
\end{equation}
where $R$ is a uniform random number between 1 and 0 \cite{bortz1975new}. 

Traps and sinks are immobile spherical objects that can either trap clusters with a certain binding energy, $E_t^\delta$, that may depend on the size of the trapped cluster, or remove them from the system. In LAKIMOCA, different traps have to be specified for different kinds of defects, such as vacancies and SIA clusters. We use traps to simulate the effect of carbon-vacancy clusters, that are able to trap SIA or vacancy clusters. Sinks are used to simulate the effect of dislocations. SIA clusters are observed by TEM to decorate dislocations \cite{hernandez2010transmission}, meaning they are not absorbed if large enough. We therefore only allow SIA clusters smaller than the core of the dislocations, \textit{i.e.} size 1--4, to be absorbed by the sinks. The number density and radius of the spherical sinks are defined to equal the sink strength of the dislocation density in the material (See \cite{nichols1978estimation}). The sink radius, $R_s^\delta$, is thus obtained as 
\begin{equation}\label{eq:sink_radius}
 R_s^\delta = \frac{\rho V }{4\pi N_s}Z^\delta-r_1^\delta
\end{equation}
where $\delta=v$ denote parameters for vacancies and $\delta=i$ parameters for SIA. $V$ is the volume of the simulation box, $N_s$ is the number of sinks. The defect radius $r^\delta$ has to be removed from the sink radius as the original sink strength expressions were derived for point-like defects. The radius $r_1^v = 4.32\cdot10^{-10}$ m is the capture radius of a single vacancy and $r_1^i = 5.17\cdot10^{-10}$ m the capture radius for a single SIA. The bias factor takes into account the strain field of defects. Since the strain field of SIA, compared to vacancies, is larger. $Z^v= 1.0$ and for SIA sinks, we tried different values.

Neutron or ion irradiation is simulated by introducing populations of vacancy and SIA clusters into the system with a certain rate per time and volume, corresponding to a certain dpa rate. 
The dpa is calculated according to the NRT standard \cite{domain2004kinetic,norgett1975proposed}
\begin{equation}
dpa = \frac{0.8 E_{MD}}{2E_D},
\end{equation} 
where $E_{MD}$ is the damage energy, the fraction of the kinetic energy of the primary knock-on atom (PKA) spectrum that is not absorbed by electronic excitation, and is well approximated by the energy of the cascades in the MD simulations. The displacement threshold energy is $E_D = 40$ eV. The cascade cluster populations are chosen randomly from a database with displacement cascade simulations \cite{stoller1996point,stoller1997primary,stoller2000statistical,
stoller2000evaluation,stoller2004secondary,stoller2000role}. The MD simulations were performed using the Finnis-Sinclair potential \cite{finnis1984simple} and using energies of
5 keV, 10 keV, 20 keV, 30 keV, 40 keV, 50 keV and 100 keV.

As anticipated in the introduction, in order to extend to higher temperatures the model that we presented in \cite{jansson2013simulation} for irradiation temperatures $<$370 K, it is necessary to allow for the presence of two types of loops. At low irradiation temperatures all SIA clusters could be assumed to be of $1/2\langle111\rangle$ type, because $\langle100\rangle$ loops are generally not observed after neutron irradiation in that temperature range \cite{zinkle2006microstructure,eyre1962direct,eyre1965electron,robertson1982low}, with only one exception for ultra-high pure Fe \cite{robertson1982low}, even though $\langle100\rangle$ loops can be produced under electron irradiation at temperatures as low as 140 K \cite{arakawa2006changes}. 

At higher temperature (550--600 K), however, the fact that $\langle100\rangle$ loops are always observed obliges the existence of this kind of defect to be somehow considered in the model \cite{malerba2011review}. The origin of this dominance of $\langle100\rangle$ SIA clusters above a certain temperature, that depends among other things on the type or irradiation (neutrons or ions), even for a given temperature, is still debated and it is not unambiguously known whether these loops are created in this orientation already in cascades (this is never seen in molecular dynamics simulations of displacement cascades) or they are the result of a subsequent transformation undergone by the clusters produced in the cascade. Here we pragmatically made the implicit assumption that the SIA clusters below the threshold for visibility can be of both kinds, whereas all visible clusters are, by default, of $\langle100\rangle$ type. This means that the migration energy, $M^i$, for SIA and the trapping energy, $E_t^i$, as 
functions of cluster size, will be different from the parameters used in \cite{jansson2013simulation}. Namely, below the visibility threshold, $N_{th}$, the energy will take an effective value, $M^i=0.2$ eV (see Sec. \ref{sec:small_sia}), that corresponds to a hypothetical weighted average between highly mobile $1/2\langle111\rangle$ loops and slower $\langle100\rangle$ loops, and similarly the trapping energy will be an effective average value; above the threshold, on the other hand, the value will be the one of  $\langle100\rangle$ loops, $M^i= 0.9$ eV \cite{osetsky2013private}. SIA clusters of sizes between 1 and 5 are considered too small to have Burgers vectors and their migration energy are the same as in \cite{jansson2013simulation}. For vacancy clusters, only the trapping energy $E_t^v$, has been changed, as will be explained later. All other parameters are unchanged and fully described in \cite{jansson2013simulation}.

% Trapping energy for SIA and VAC and C-V complexes
Another effect of temperature that must be taken into account is that the carbon-vacancy complexes stable at low and high temperature are not the same (this aspect was already discussed in \cite{jansson2013simulation} and applied for the simulation of the post-irradiation annealing).
Simulations of small C-V complexes in \cite{jansson2013simulation} revealed indeed that the two dominating C-V complexes at temperatures above 450 K are C and C$_2$V. Molecular dynamics (MD) studies shows that C binds with the edge of a $\langle100\rangle$-SIA loop with an energy of 1.1 eV, but is repulsed inside the glide prism of the loop. C$_2$V complexes were calculated to bind to a 61-$\langle100\rangle$ SIA with an energy of 0.6 eV \cite{anento2013unpublished}. As in \cite{jansson2013simulation}, we simulate the effect of these complexes using generic spherical traps for SIA and vacancies. The estimated amount of C in the the reference material used in the experiment was $\sim$65 appm. If we also include the amount of N atoms, that are assumed to have very similar properties to C atoms, we get a total amount of 134 appm. A priori, other interstitial impurities with similar effects might be present as well, \textit{e.g.} oxygen or hydrogen. Here, and in what follows, they will be formally all confused 
with C, which is the main one. We then used two kinds of traps for SIA clusters, corresponding to C atoms and C$_2$V complexes. Since the latter complex contains twice as many C atoms as the former, the total density of traps will be less than the 
amount of carbon in the system. We used 66 appm of C traps ($E_{t1}^v$) and 34 appm of C$_2$V traps ($E_{t2}^v$); together they correspond to 100 appm and therefore to 134 appm C. The 2:1 ratio of the densities of the two SIA traps were taken from the results of the annealing simulation with carbon explicit in \cite{jansson2013simulation}. We used consistently 100 appm of traps for vacancies ($E_t^v$). The values are listed in Table \ref{table:M_i}. In this table the only differences with respect to the parameters in \cite{jansson2013simulation} concerns clusters above size 6: up to the visibility threshold, migration and trapping energy are effective values that result from an iteration; above the threshold, the trapping energy for $\langle100\rangle$ loops is taken from \cite{anento2013unpublished}, while the migration energy was chosen based on indications from work \cite{osetsky2013private} performed using a recently developed on-the-fly KMC scheme \cite{xu2011simulating}. The vacancy trapping energy was 
also the result of iterations, as is shown in the section on results. The capture radius was 5.0 Å for all traps, as was used in \cite{jansson2013simulation,jansson2014okmc}.

% Experimental set-up
We used a simulation box size of $450\times520\times600a_0^3$, where $a_0 = 2.87\cdot10^{-10}$ m is the lattice parameter for $\alpha$-Fe. Periodic boundary conditions were applied in all three directions. Simulation temperature, grain size, dislocation density, and dpa rate were dictated by the reference experiments in the REVE campaign \cite{lambrecht2008influence,lambrecht2009phd,lambrecht2009positron,bergner2010comparative,hernandez2010transmission,meslin2010characterization} (See introduction). 

\section{Results}\label{sec:results}

\subsection{The cluster density and mean size evolution compared to the reference experiment} 

In this section we present the results obtained with the parameter choice that gives globally the best results as compared to experiments. For this we used $N_{th} = 90$, the visibility size threshold for SIA clusters; $E_t^v = 0.2$ eV, the trapping energy of vacancy clusters; and $Z^i = 3.0$, the bias between vacancy and SIA sink radii. The migration energy used is $M^i = 0.2$ eV for SIA clusters of size 6--$N_{th}$ and 0.9 eV above $N_{th}$. These values will be discussed in more detail in the sections to follow.

The evolution of the vacancy clusters can be compared with results from two different experimental techniques, PAS and SANS, in terms of density and mean cluster size. However, the statistical analysis has to take into account that the two techniques are sensitive to different parts of the vacancy cluster population. PAS is especially suitable for small vacancy clusters, being sensitive even to the presence of single vacancies. However, it provides correct information about the size only for vacancy clusters up to a diameter of $\sim$1 nm ($N^v\sim 50$), above which all clusters sizes are indistinguishable. SANS, on the other hand, can not detect clusters smaller than $\sim$1 nm in diameter. 

\begin{figure}
 \centering
  \includegraphics[width=\columnwidth]{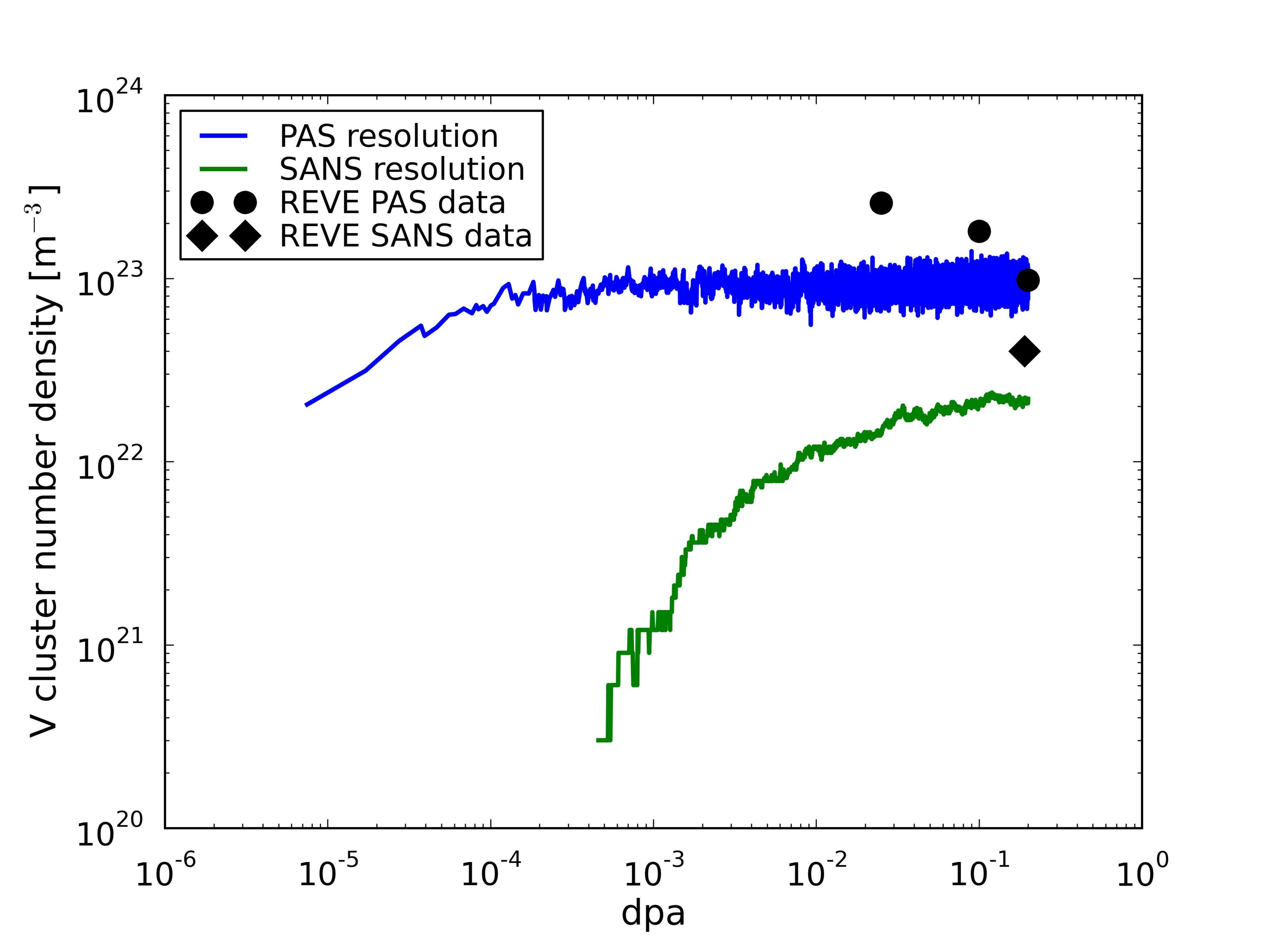}
    \caption{Density of vacancy clusters of sizes observable by PAS and SANS, respectively, versus dpa. The data are compared with PAS \cite{lambrecht2009phd,meslin2010characterization} and SANS data \cite{bergner2010comparative} from the REVE campaign.}
  \label{R0110-Z30-Ev04-60_298967.pdf}
\end{figure}

\begin{figure}
 \centering
  \includegraphics[width=\columnwidth]{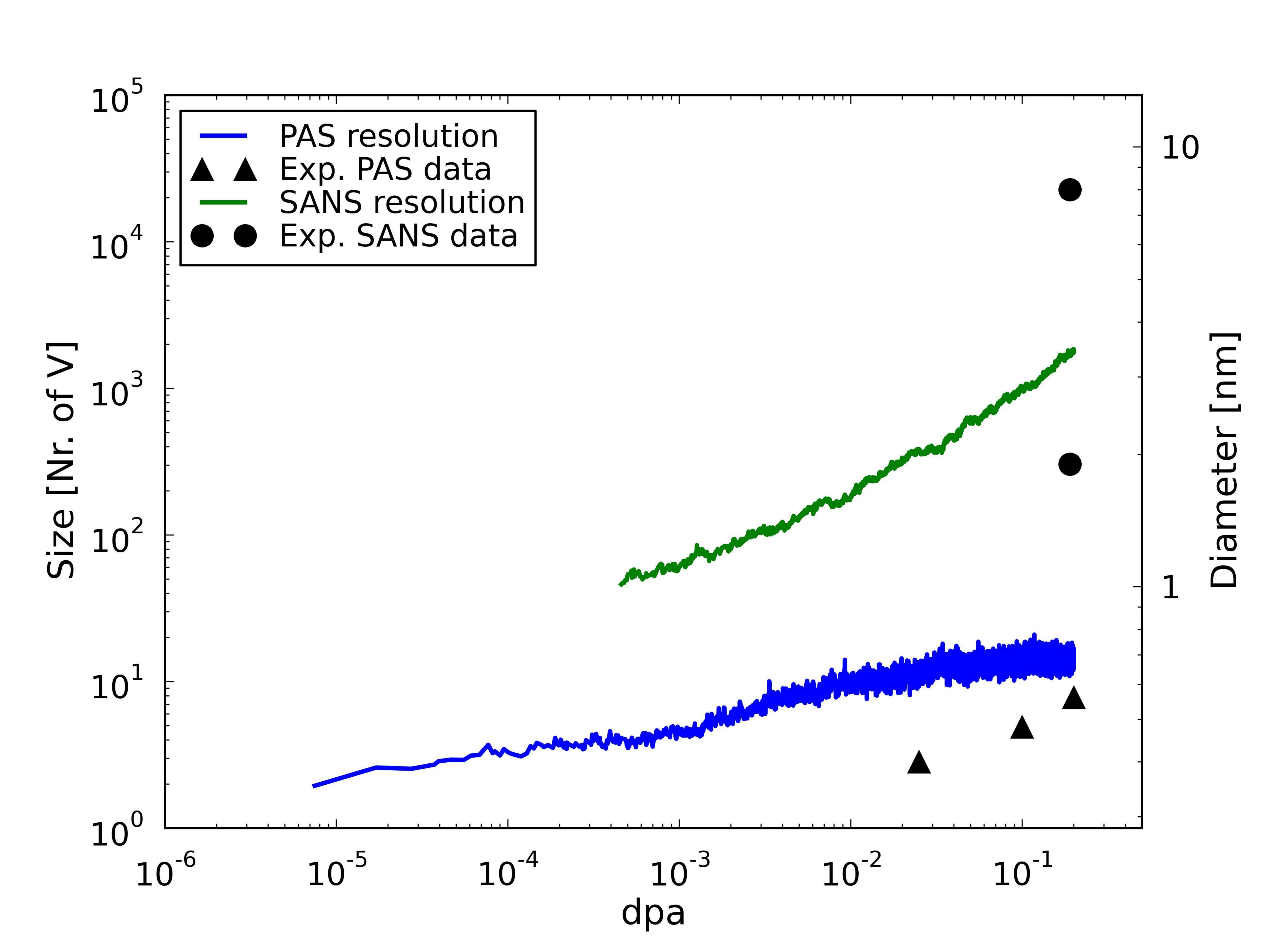}
    \caption{Vacancy cluster mean size versus dpa with PAS and SANS resolution. The experimental PAS data (triangles) are from \cite{lambrecht2009phd,meslin2010characterization}. The SANS data (bullets) shows the two peaks of the size distribution at 0.2 dpa from \cite{bergner2010comparative}. The lower SANS data point is the major peak.}
  \label{R20130417-35_464423_vac_mean_size_evolution.pdf}
\end{figure}

\begin{figure}
 \centering
  \includegraphics[width=\columnwidth]{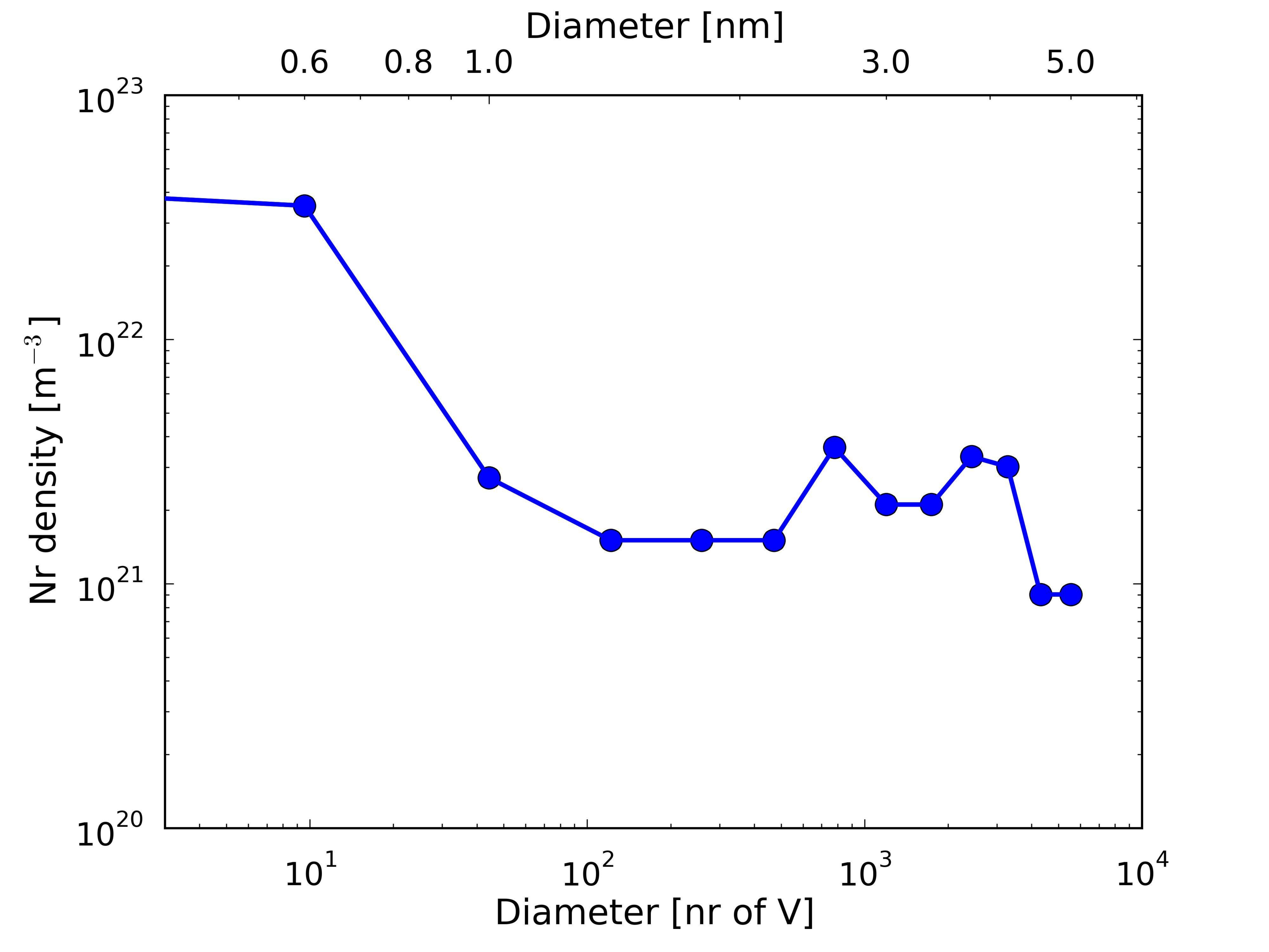}
    \caption{Vacancy cluster size distribution at 0.2 dpa. The bin size is 0.4 nm.}
  \label{monica_V_size_distribution.pdf}
\end{figure}

In Fig. \ref{R0110-Z30-Ev04-60_298967.pdf} the number density of vacancy clusters of all sizes is shown as a function of dose. The density at 0.2 dpa is $10^{23}$ m$^{-3}$, which coincides with the experimental PAS value \cite{lambrecht2009phd,meslin2010characterization}. The vacancy cluster mean size versus dpa is shown in Fig. \ref{R20130417-35_464423_vac_mean_size_evolution.pdf} and the size distribution at 0.2 dpa is shown in Fig. \ref{monica_V_size_distribution.pdf}. The over-all mean size is $\sim$2 nm. If one considers that PAS does not distinguish a 50 vacancy cluster from bigger ones, the mean size is $\sim$0.7 nm, which is only slightly larger than the experimental PAS value of 0.6 nm
\cite{lambrecht2009phd,meslin2010characterization}.

If we now consider only the cluster population detectable by SANS, the mean size is $\sim$3 nm and the density $2.20\cdot10^{22}$ m$^{-2}$. However, in the actual experiment, SANS sees in fact a bimodal size distribution, with one peak at 1.9 nm (cluster diameter) and another one at 8 nm \cite{bergner2010comparative}. TEM studies have observed voids with a mean size of $12\pm0.4$ nm and an estimated density of $1.2\cdot10^{20}$ m$^{-3}$. The TEM peak is likely to be the same as the second peak seen by SANS. The density corresponding to this peak with large 8--12 nm clusters is, however, too low to be seen in our simulations. The mean size obtained for clusters larger than 1 nm by the simulations, 3.4 nm, thus corresponds well to the SANS and TEM data, being slightly higher than the main peak value, 1.9 nm, but still between the two peaks. The density of the clusters in the first peak of the SANS distribution is estimated to be $4\cdot10^{22}$ m$^{-3}$ \cite{bergner2010comparative}, slightly above our 
simulated values.

\begin{figure}
 \centering
  \includegraphics[width=\columnwidth]{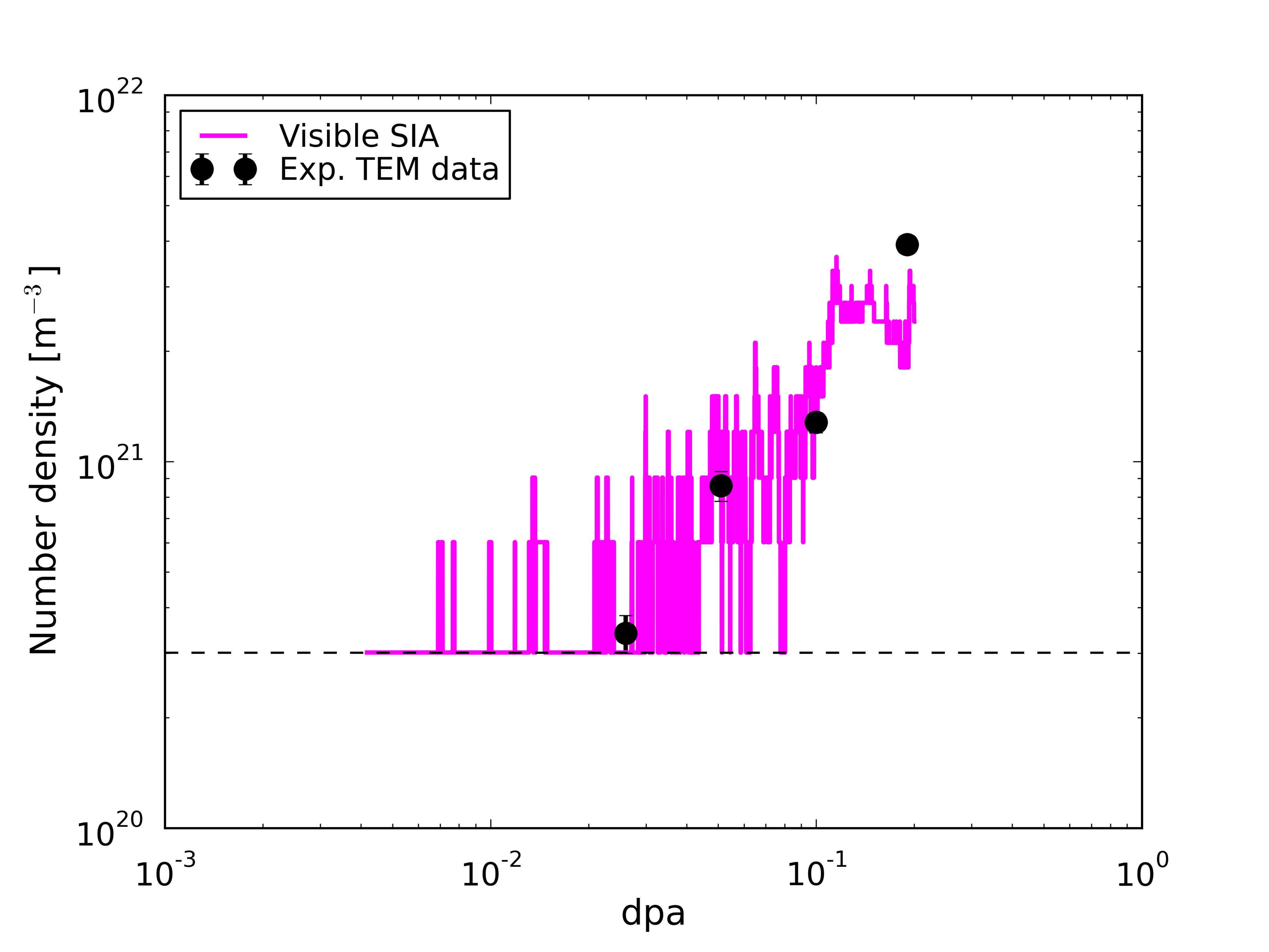}
    \caption{Visible SIA cluster density evolution versus dpa. The experimental TEM data are from \cite{hernandez2010transmission}. The dotted line correspond to one cluster in the simulated volume.}
  \label{R20130417-35_464423_visible_SIA.pdf}
\end{figure}

\begin{figure}
 \centering
  \includegraphics[width=\columnwidth]{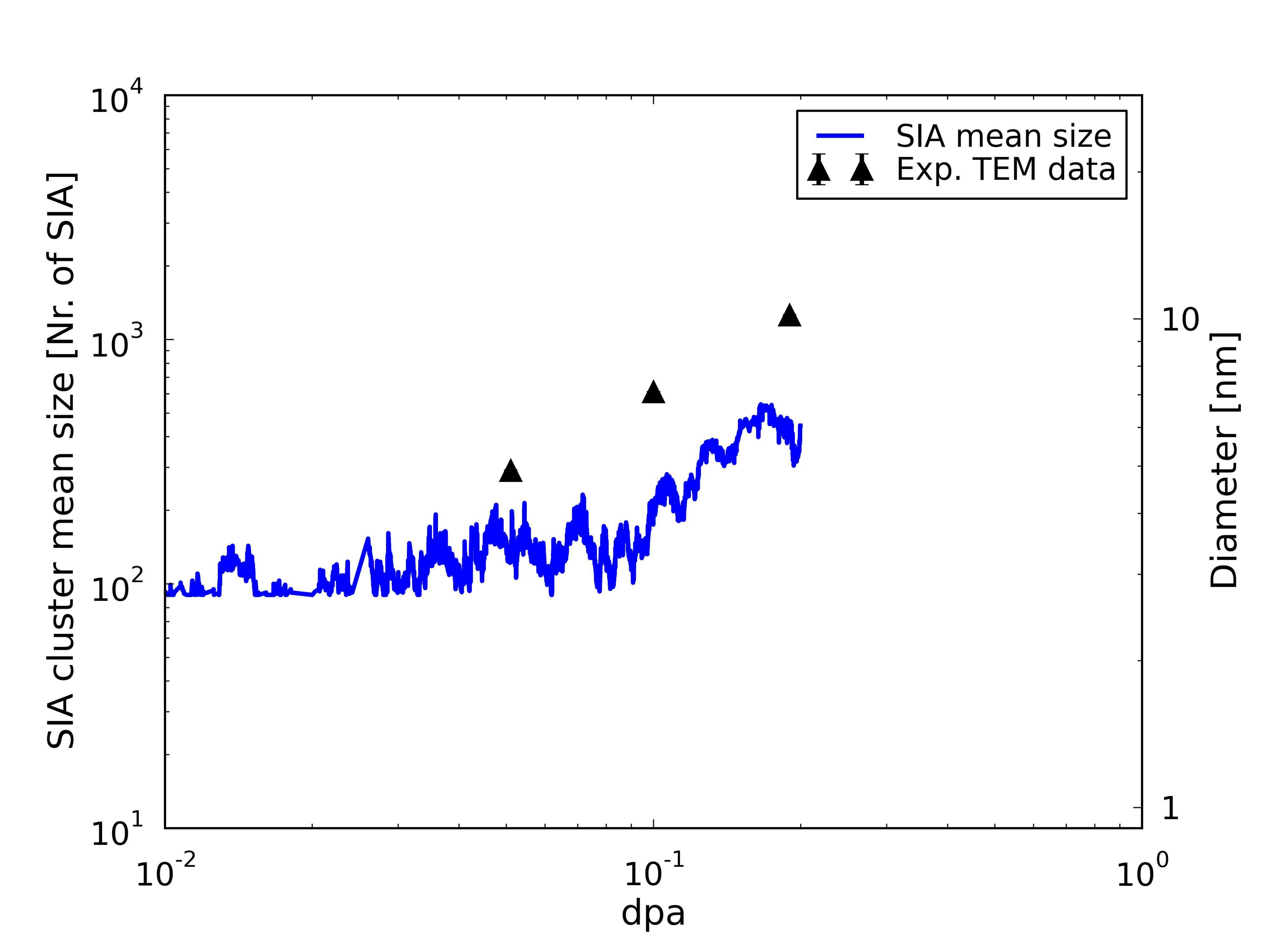}
    \caption{Visible SIA cluster mean cluster size evolution versus dpa. The experimental TEM data are from \cite{hernandez2010transmission}.}
  \label{R20130417-35_464423__best_case_SIA_mean_size_evolution.pdf}
\end{figure}

The best cases for the visible SIA cluster evolution in terms of number density and cluster size are shown in Fig. \ref{R20130417-35_464423_visible_SIA.pdf} and Fig. \ref{R20130417-35_464423__best_case_SIA_mean_size_evolution.pdf}, respectively. The density shows good agreement with the experimental data from \cite{hernandez2010transmission}, whereas the mean cluster sizes are a bit underestimated by the model.

\subsection{Effect of the choice of the visibility threshold}

\begin{figure}
 \centering
  \includegraphics[width=\columnwidth]{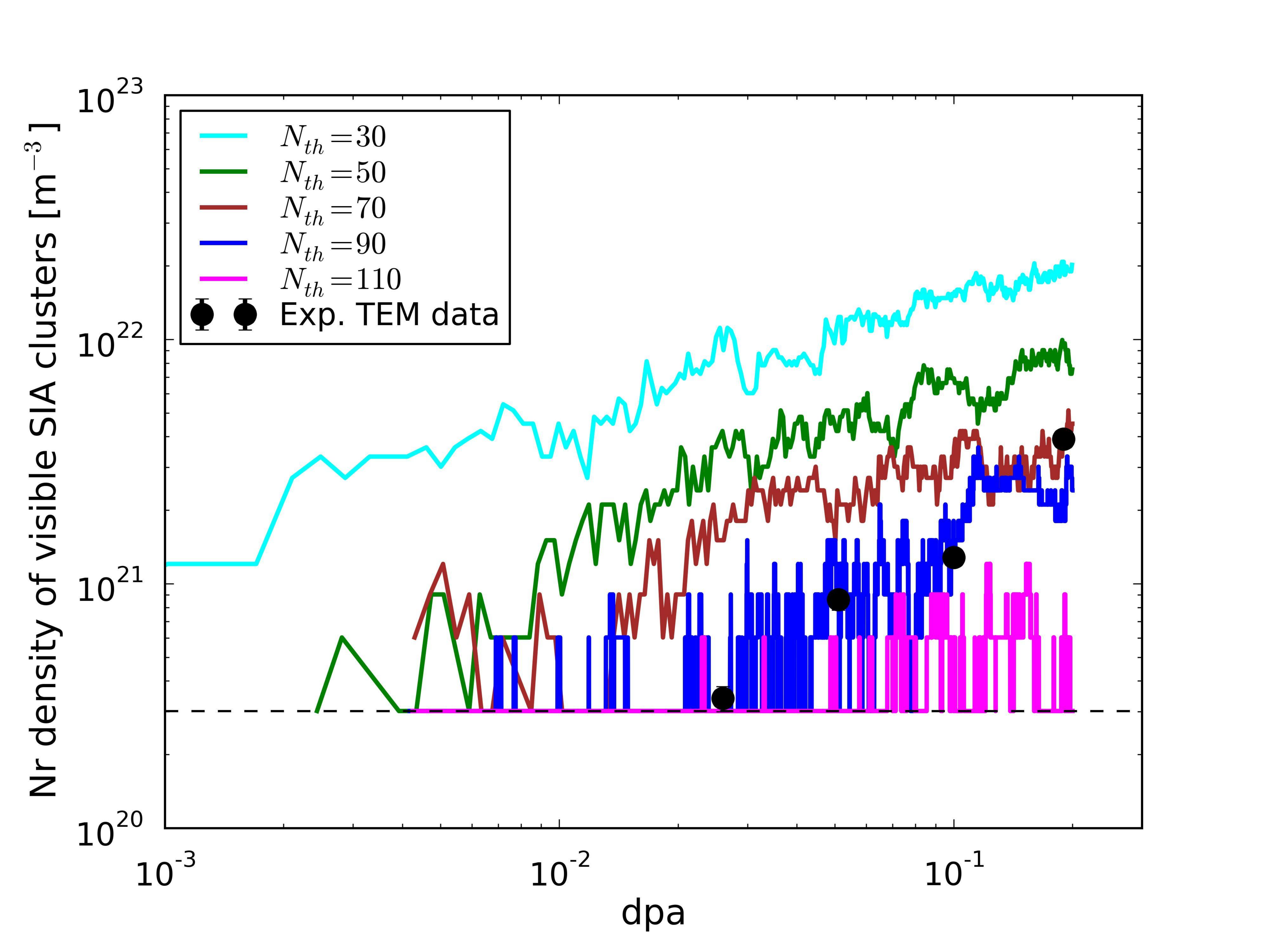}
    \caption{Visible SIA cluster density versus dpa for different values of the visibility threshold, $N_{th}$. The experimental TEM data are from \cite{hernandez2010transmission}. The dotted line corresponds to one cluster in the simulated volume.}
  \label{monica_Nth.pdf}
\end{figure}

\begin{figure}
 \centering
  \includegraphics[width=\columnwidth]{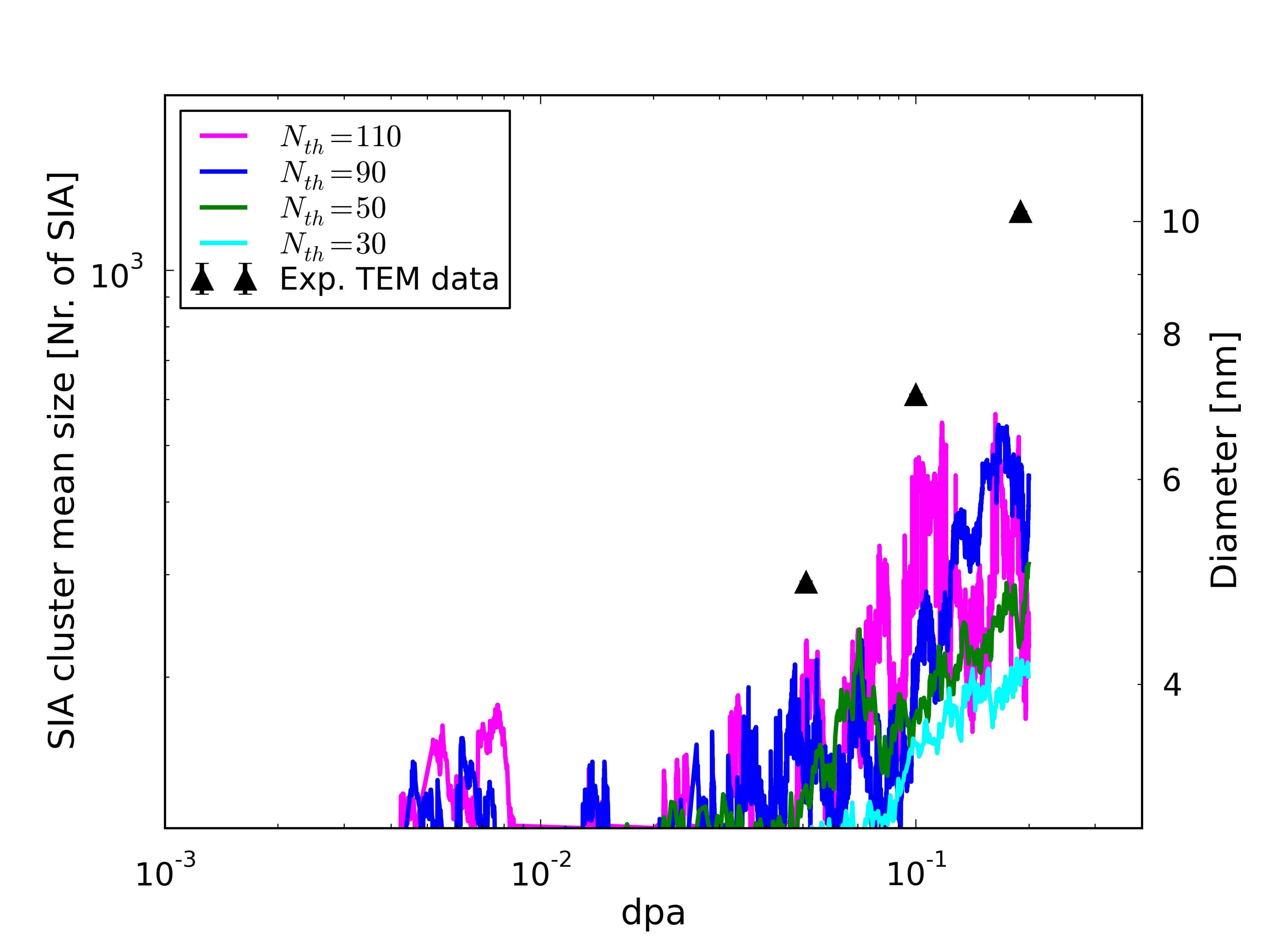}
    \caption{The SIA mean cluster size versus dpa for different values of the threshold, $N_{th}$. The experimental TEM data are from \cite{hernandez2010transmission}.}
  \label{R0110-Z30-Ev04-60_298967_SIA_mean_size.pdf}
\end{figure}

For the SIA migration energy, $M^i$, for different sizes, as presented in Table \ref{table:M_i}, there are two values that may change and need to be established: the "effective" migration energy for the "mixture" of $\langle111\rangle$ and $\langle100\rangle$ clusters below the threshold for visibility and, to a certain extent, the latter threshold as well. Moreover, the value of 0.9 eV for $\langle100\rangle$ loops is based only on preliminary calculations. We therefore did trials with different values for $M^i$ below, as discussed in the next section, and above size $N_{th}$ (but above size $N^i>5$). The values 0.2 eV and 0.9 eV gave the best results. The effect of choosing different threshold sizes for visibility, $N_{th}$, was also studied. A few cases are shown in Fig. \ref{monica_Nth.pdf} for the visible SIA cluster density. Obviously, a higher $N_{th}$ gives 
lower visible SIA cluster densities, while the SIA cluster growth seems not to be significantly affected. Using $N_{th}=90$, good agreement is obtained with the experimental data \cite{hernandez2010transmission} both in terms of density and mean SIA sizes, the latter being shown for different $N_{th}$ in Fig. \ref{R0110-Z30-Ev04-60_298967_SIA_mean_size.pdf}. No significant impact on the vacancy cluster densities and cluster mean sizes were observed when varying the $N_{th}$ parameter. This is likely because the vacancy trapping energy, $E_t^v$, which is the important parameter for the vacancy clusters, is not dependent on the $N_{th}$ parameter.
\begin{table}
  \centering
   \caption{Overview of the parameters for irradiation temperatures above 370 K, as used in this paper. All are a function of the cluster size, $N^\delta$, where $\delta$ = $i$ for SIA clusters and $v$ for vacancy clusters. $M^i$ is the migration energy for SIA clusters. The values for size 1 to 5 are the same as in \cite{jansson2013simulation}. $E_{t1}^i$ and $E_{t2}^i$ are the SIA trapping energies representing C and C$_2$V complexes, respectively. $E_t^v$ are the trapping energy for vacancy clusters. It is worth noting that the $E_{t2}^i$ energy is only valid up to $\sim$700 K, as the C$_2$V complexes will dissolve to C atoms and vacancies at this temperature.}
\label{table:M_i}
\begin{tabular*}{\columnwidth}{@{\extracolsep{\fill}} l l l l l}
\toprule
$N^\delta$	& $M^i$	& $E_{t1}^i$ 	& $E_{t2}^i$  	& $E_t^v$\\
	& [eV]	& [eV]       	& [eV]		& [eV]   \\
\midrule
1 	& 0.31 	& 0.17 \cite{becquart2011p60}			& 0.6		& 0.65 \cite{becquart2011p60}	\\
2 	& 0.42 	& 0.28 \cite{becquart2011p60}			& 0.6		& 1.01 \cite{becquart2011p60}	\\	
3 	& 0.42 	& 0.36 \cite{becquart2011p60}			& 0.6		& 0.93 \cite{becquart2011p60}	\\
4 	& 0.80 	& 0.34 \cite{becquart2011p60}			& 0.6		& 0.96 \cite{becquart2011p60}	\\
5 	& 0.10 	& 0.60 \cite{jansson2013simulation}		& 1.2		& 1.23 \cite{becquart2011p60}	\\
6 	& 0.20 	& 0.60 \cite{jansson2013simulation}		& 1.2		& 1.20 \cite{becquart2011p60}	\\
7--$N_{th}$ & 0.20 & 0.60 	& 1.2		& 0.4 	\\
$N_{th}<$   & 0.9  & 1.1     	& 0.6		& 0.4 	\\
\bottomrule
\end{tabular*}
\end{table}

\subsection{Effect of varying the migration energy for invisible SIA clusters}\label{sec:small_sia}

\begin{figure}
 \centering
  \includegraphics[width=\columnwidth]{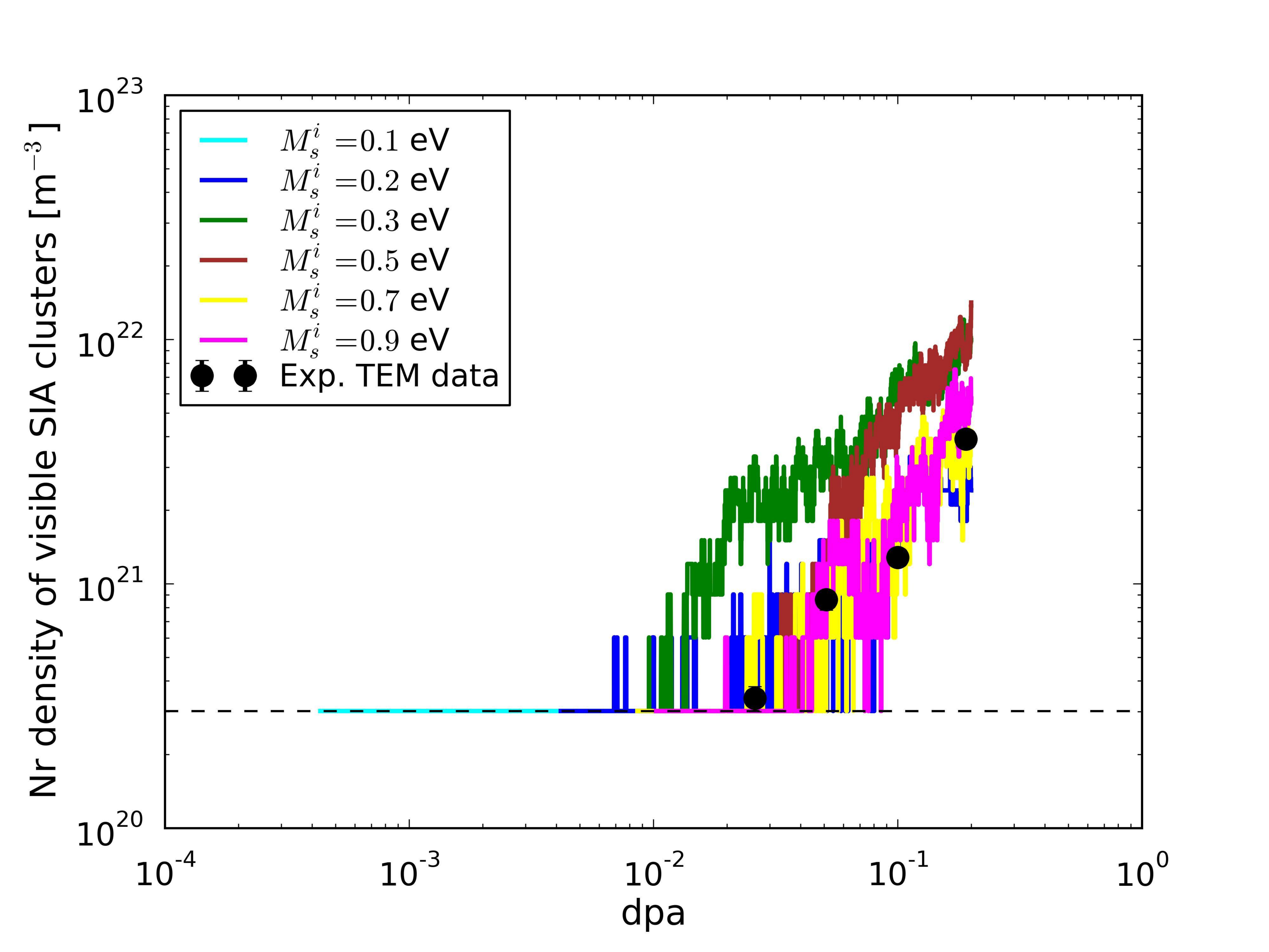}
    \caption{Visible SIA cluster density versus dpa for different migration energies $M^i_s$ for SIA clusters of size $6\leq N^i < N_{th}$. The experimental TEM data are from \cite{hernandez2010transmission}. The dotted line corresponds to one cluster in the simulated volume.}
  \label{monica_Ms.pdf}
\end{figure}

\begin{figure}
 \centering
  \includegraphics[width=\columnwidth]{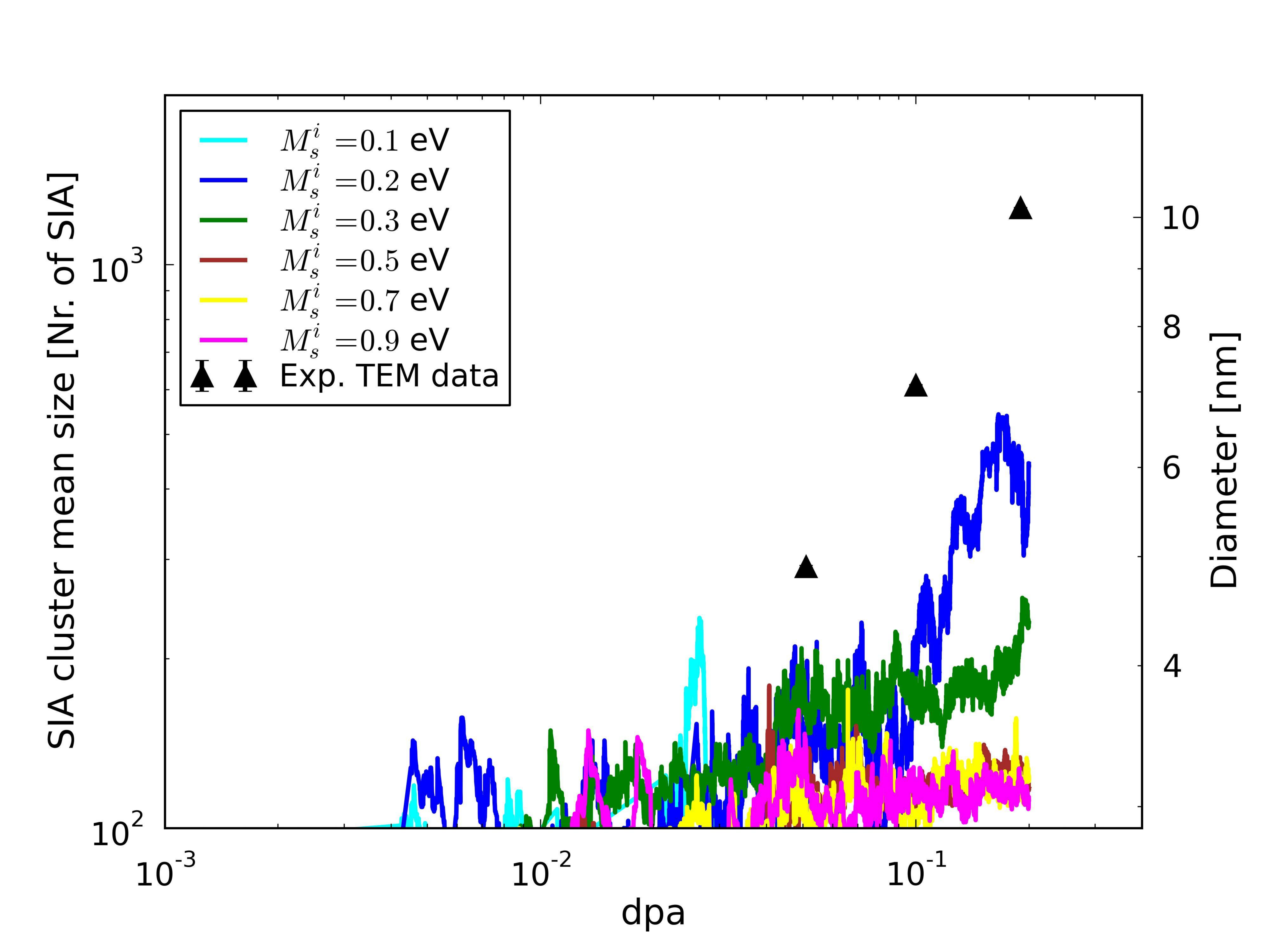}
    \caption{The SIA mean cluster size versus dpa for different migration energies $M^i_s$ for SIA clusters of size $6\leq N^i < N_{th}$. The experimental TEM data are from \cite{hernandez2010transmission}.}
  \label{monica_Ms_SIA_mean_size_evolution.pdf}
\end{figure}

Figs. \ref{monica_Ms.pdf} and \ref{monica_Ms_SIA_mean_size_evolution.pdf} shows the visible SIA cluster density and mean size, respectively, for different migration energy of SIA clusters of small sizes, $6\leq N^i < N_{th}$, \textit{i.e.} invisible cluster sizes. For sizes above $N_{th}$, $M^i=0.9$ eV, as described above and in Table \ref{table:M_i}. The cluster density does not vary significantly by varying the migration energy between 0.2 eV and 0.9 eV for invisible clusters. With 0.1 eV, no SIA clusters grow above the visibility threshold size, $N_{th}$. With a slightly higher migration energy, the SIA clusters are slowed down enough for nucleation of larger clusters to occur. Using $M^i=0.2$ eV, gives the largest mean SIA cluster size and also the best agreement with the experimental data. The effect of the $M^i$ for invisible SIA clusters on the vacancy density and mean cluster size is minimal. Over all, $M^i=0.2$ eV gives the best result. 

\subsection{Effect of varying the vacancy cluster trapping energy}
\begin{figure}
 \centering
  \includegraphics[width=\columnwidth]{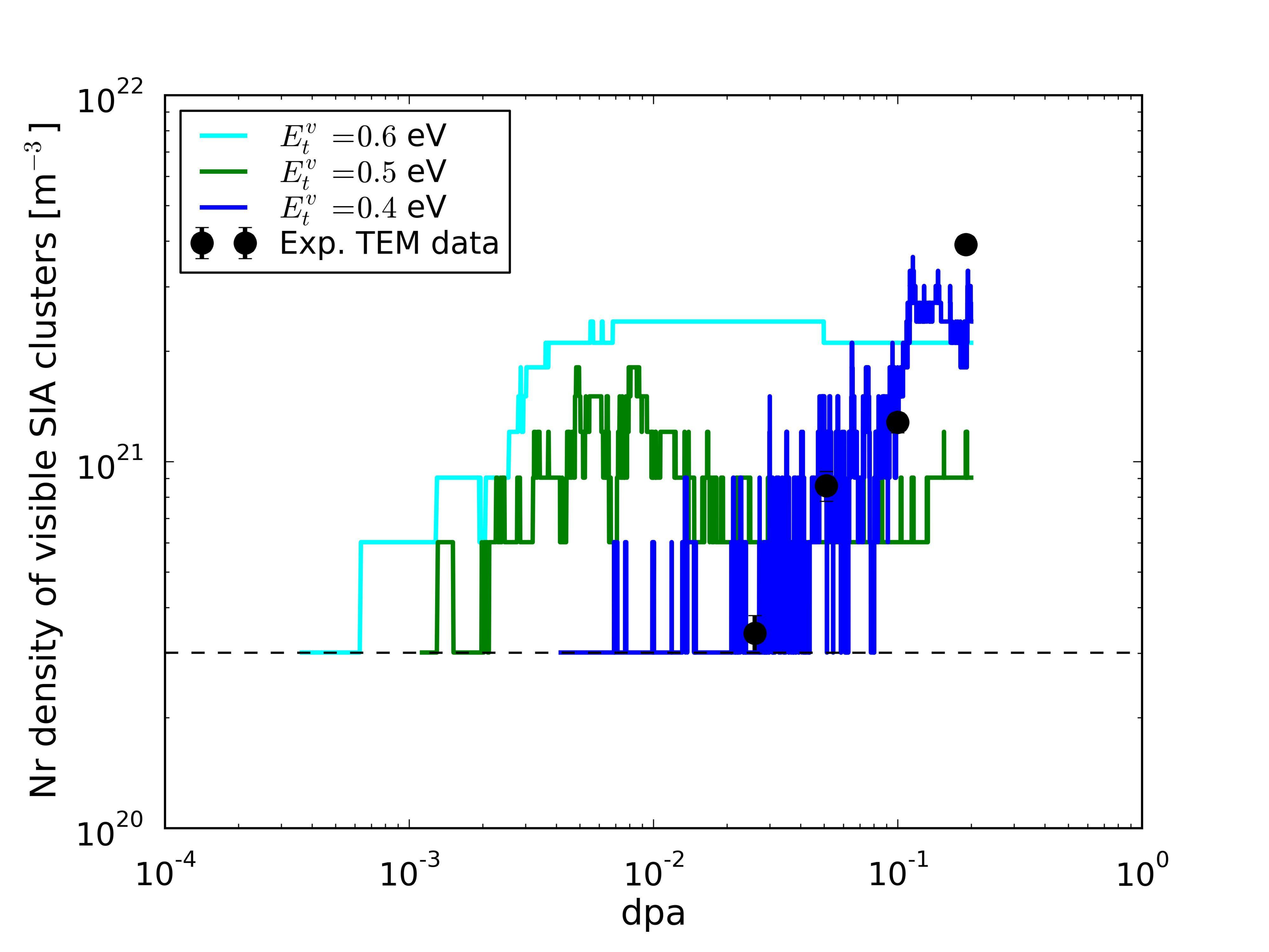}
    \caption{Visible SIA cluster density versus dpa for different values of the trapping energy for vacancy clusters above size 6. The experimental TEM data are from \cite{hernandez2010transmission}. The dotted line corresponds to one cluster in the simulated system.}
  \label{monica_Etvac.pdf}
\end{figure}

\begin{figure}
 \centering
  \includegraphics[width=\columnwidth]{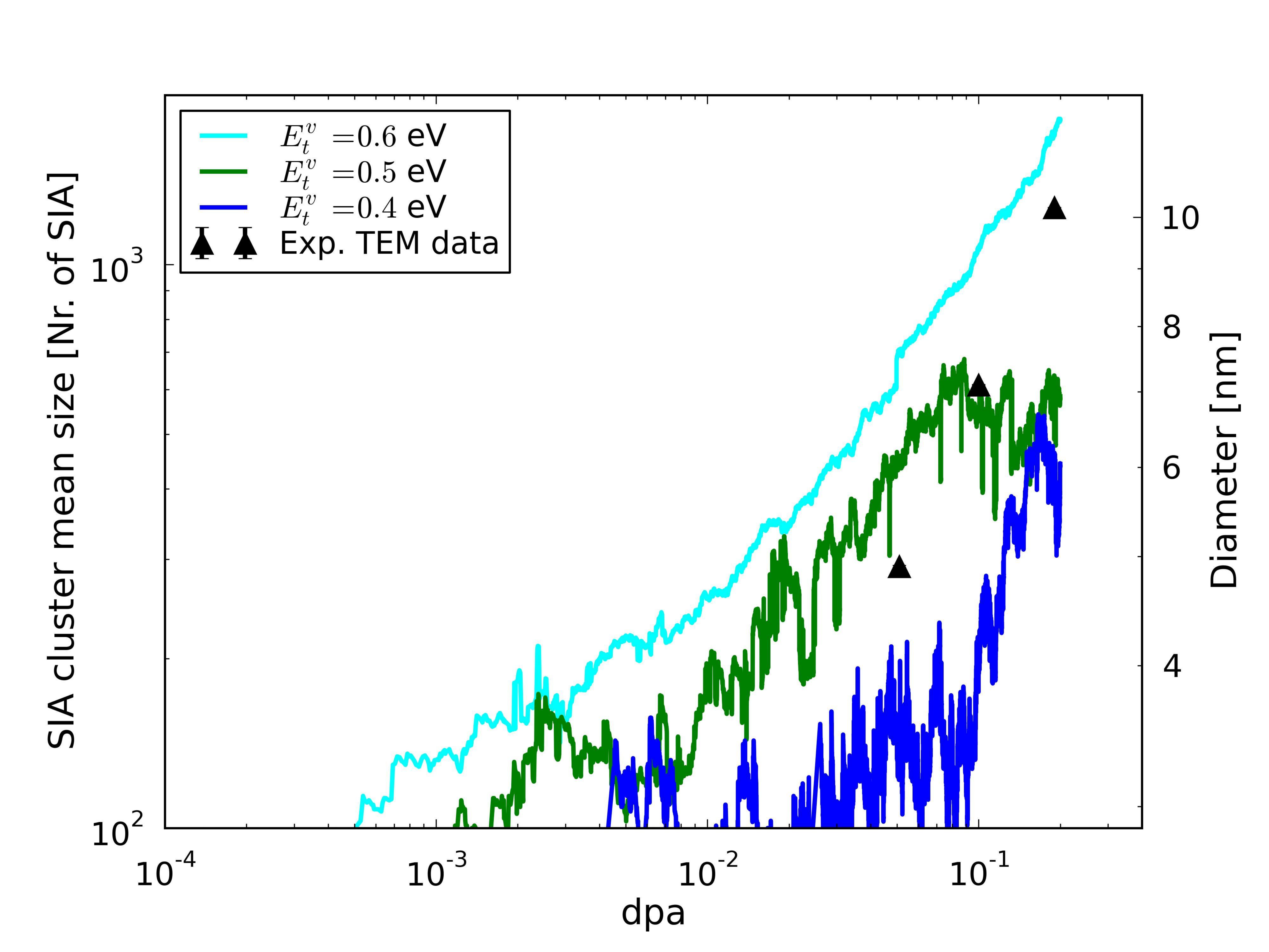}
    \caption{SIA cluster mean size versus dpa for different values of the trapping energy for vacancy clusters above size 6. The experimental TEM data are from \cite{hernandez2010transmission}.}
  \label{monica_Etvac_SIA_mean_size_evolution.pdf}
\end{figure}

\begin{figure}
 \centering
  \includegraphics[width=\columnwidth]{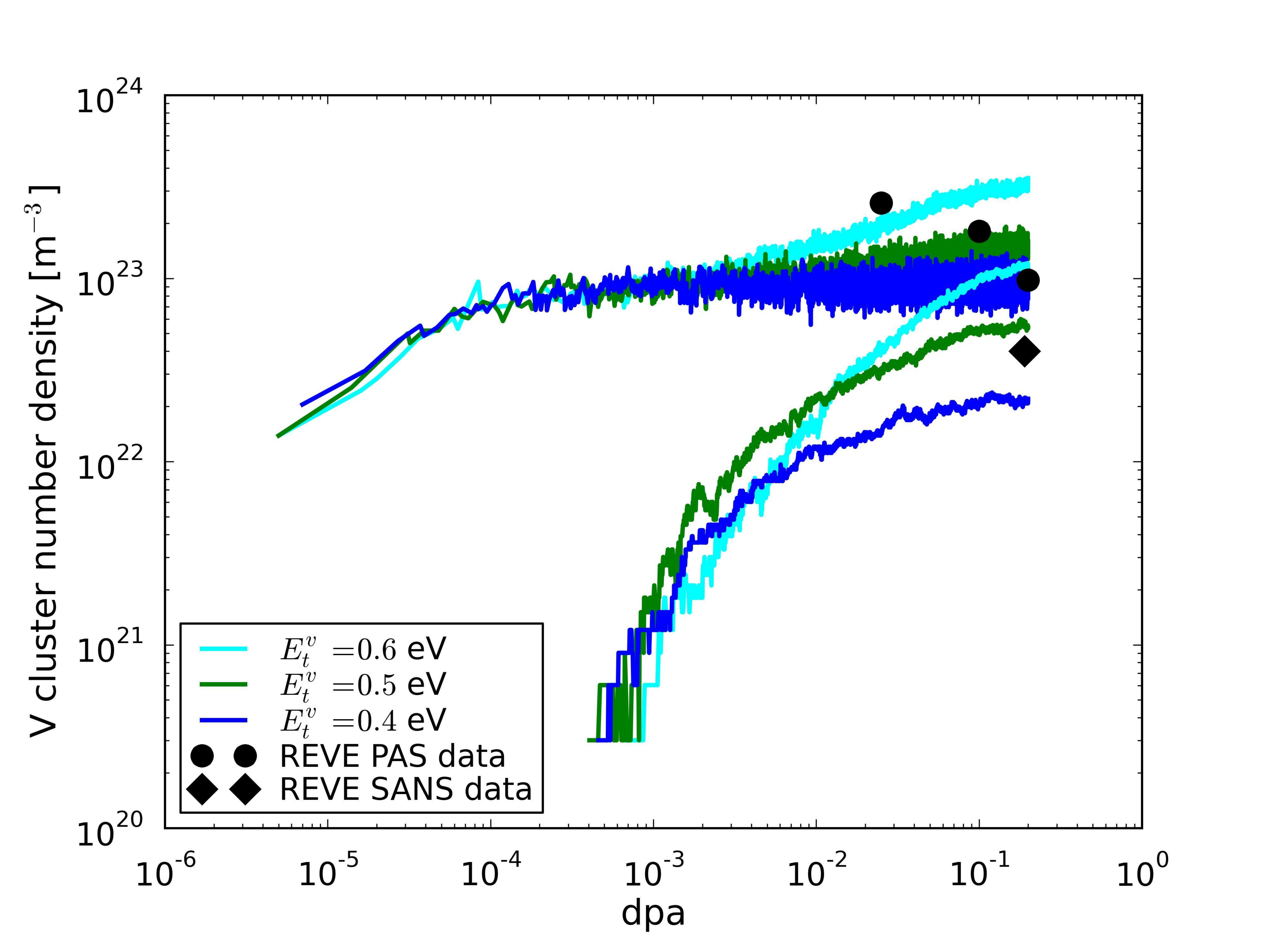}
    \caption{Vacancy cluster densities versus dpa for different $E_t^v$ values. Higher densities are calculated with the PAS size resolution; lower densities with the same colour are calculated with the SANS size resolution. Results are compared with experimental PAS \cite{lambrecht2009phd,meslin2010characterization} and SANS data \cite{bergner2010comparative} from the REVE campaign.}
  \label{monica_Etvac_vac_density_evolution.jp}
\end{figure}

\begin{figure}
 \centering
  \includegraphics[width=\columnwidth]{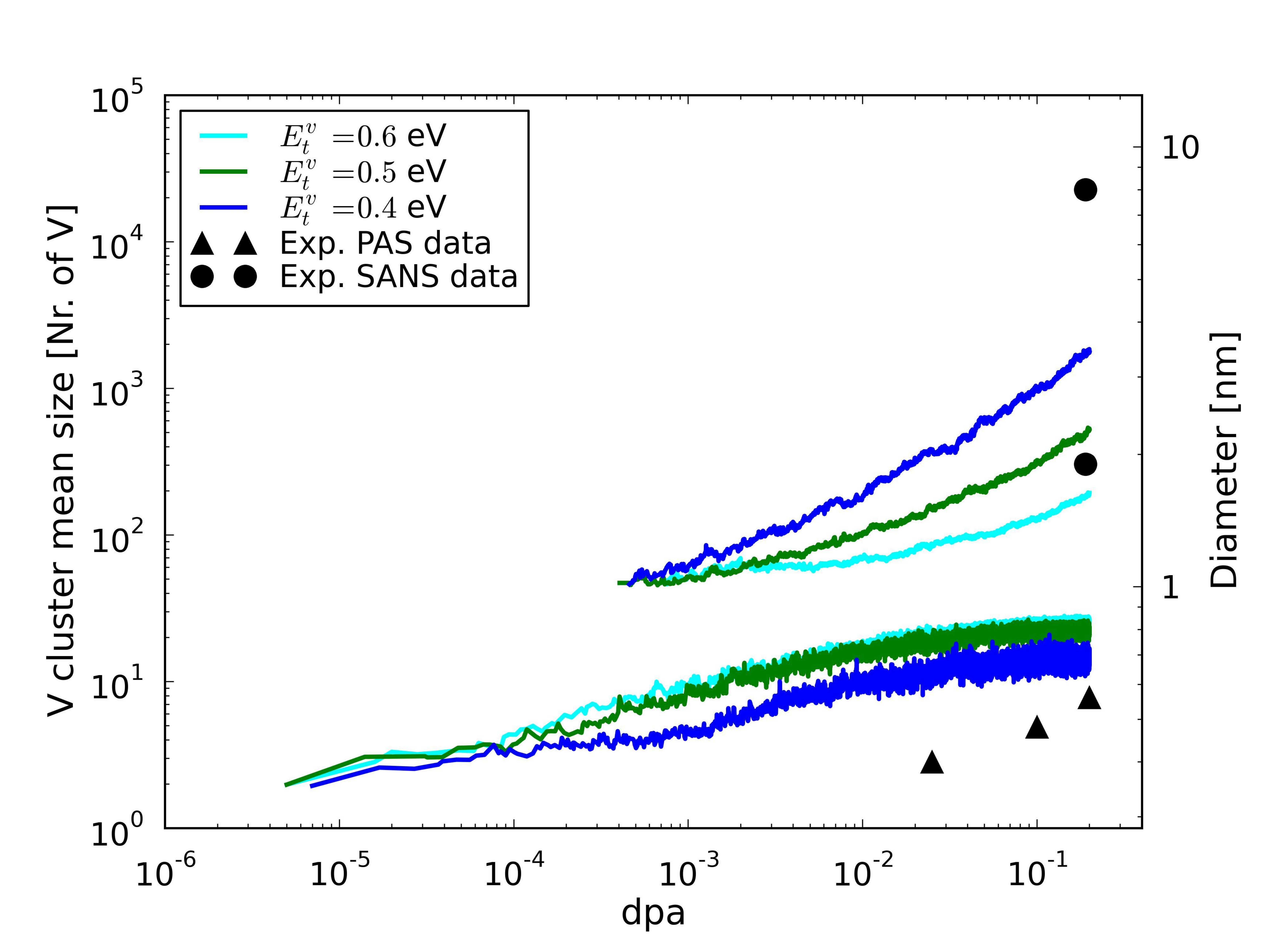}
    \caption{Vacancy cluster mean sizes versus dpa for different $E_t^v$ values. Larger mean sizes are calculated using the SANS size resolution; smaller mean sizes are calculated using the PAS resolution. Results are compared with experimental PAS \cite{lambrecht2009phd,meslin2010characterization} and SANS data \cite{bergner2010comparative} from the REVE campaign. SANS points corresponds to the two peaks in the observed size distribution: the lower point is the major peak.}
  \label{monica_Etvac_vac_mean_size_evolution.pdf}
\end{figure}

Vacancies are more mobile at 563 K than at 343 K, which was the irradiation temperature in our previous work \cite{jansson2013simulation}. Thus, vacancy traps play a more important role in this situation. In \cite{jansson2013simulation}, only V clusters of sizes $N^v \leq 6$ were trapped, essentially because precise data were available only up to that size and above it the migration energy was already so large that cluster migration would become a very rare event. In this study we tried different trapping energy for vacancy clusters above size $N^v = 6$. This trapping effect would correspond to the formation of complexes with C or other interstitial impurities. The results are shown in Fig. \ref{monica_Etvac.pdf} and \ref{monica_Etvac_SIA_mean_size_evolution.pdf} for the visible SIA cluster density and mean size evolution, respectively. For the vacancy clusters, the density and mean size evolutions are shown in Fig. \ref{monica_Etvac_vac_density_evolution.jp} and \ref{monica_Etvac_vac_mean_size_evolution.pdf}
, respectively. It can be seen that this parameter has a quite large effect, when varied. A higher $E_t^v$ increases both the cluster densities and the cluster mean sizes for both SIA and vacancy clusters. The exception to this trend is when considering only the vacancy clusters with the larger SANS resolution, whose mean size actually decreases with increased $E_t^v$. This effect is likely to be due to the fact that more strongly trapped vacancy clusters lead to a decrease of the number of recombinations and clustering of vacancy clusters, thereby decreasing the mean size of the large vacancy clusters visible by SANS, but increasing both size and density of the smaller vacancy clusters, seen by PAS. Less recombination leads to more SIA clusters and faster SIA cluster growth.

Taking all data for both SIA clusters and vacancies into account, the best fit is given by $E_t^v = 0.4$ eV, because it gives good agreement for the cluster densities and fair agreement for the cluster mean sizes. If higher values are chosen, the mean size is closer to experiment but the density of SIA clusters is totally off, because it builds up and saturates too early. The trapping energy for all vacancy cluster sizes are listed in Table \ref{table:M_i}. We confirmed by repeating the irradiation experiment at $T=343$ K, reported in \cite{jansson2013simulation}, that these $E_t^v$ values do not change the results at lower temperature.

\subsection{Determination of the bias between SIA and vacancy sink radii}
\begin{figure}
 \centering
  \includegraphics[width=\columnwidth]{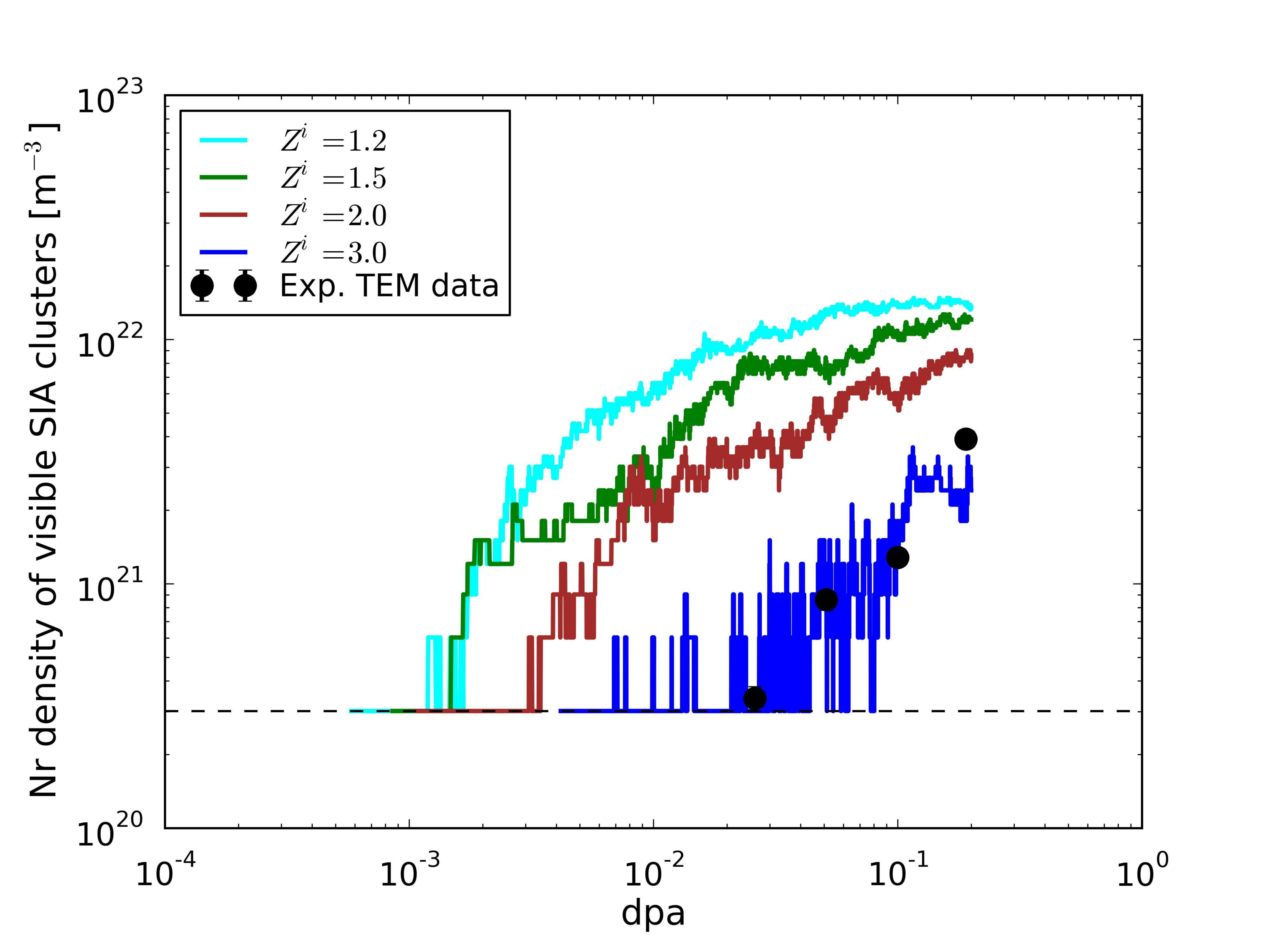}
    \caption{Visible SIA cluster density versus dpa for two different values of the SIA bias factor for sinks, $Z^i$. The experimental TEM data are from \cite{hernandez2010transmission}.  The dotted line corresponds to one cluster in the simulated volume.}
  \label{monica_Z.pdf}
\end{figure}

\begin{figure}
 \centering
  \includegraphics[width=\columnwidth]{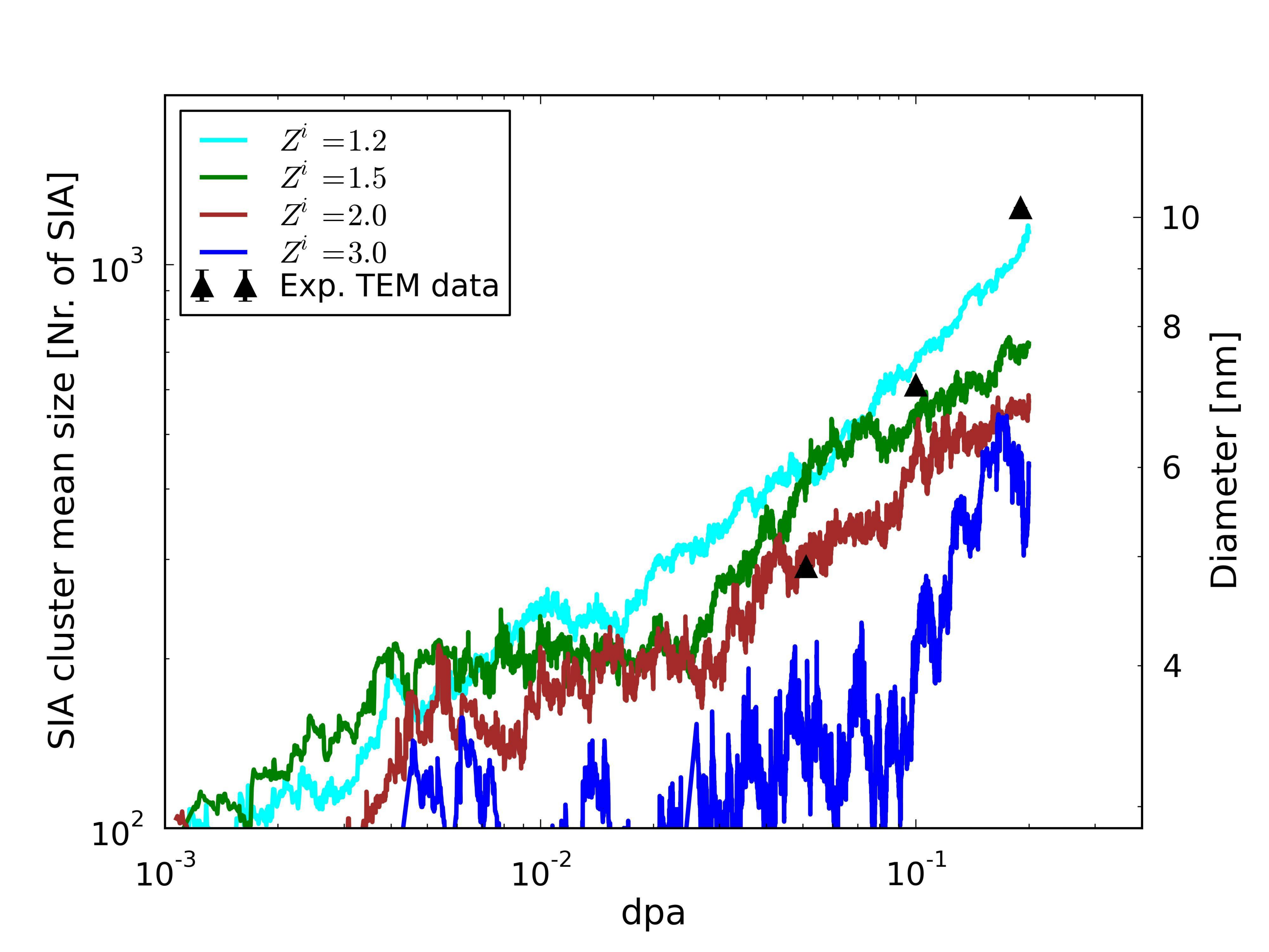}
    \caption{SIA mean cluster size versus dpa. The experimental TEM data are from \cite{hernandez2010transmission}.}
  \label{monica_Z_SIA_mean_size_evolution.pdf}
\end{figure}

\begin{figure}
 \centering
  \includegraphics[width=\columnwidth]{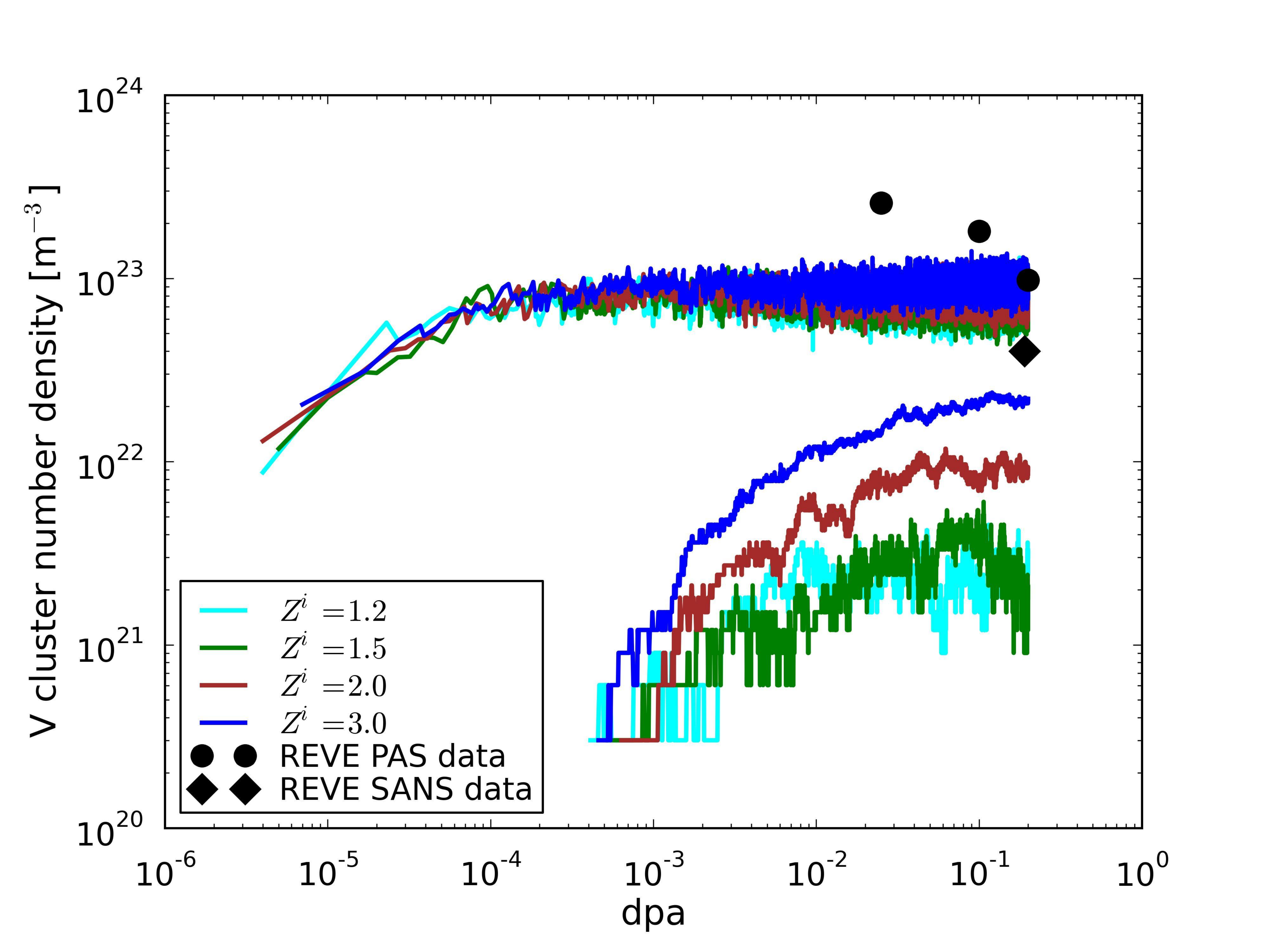}
    \caption{Vacancy cluster densities versus dpa for different $Z^i$ values. The higher densities are calculated with the PAS size resolution, whereas the lower densities with the same colour are calculated with the SANS size resolution. The data are compared with experimental PAS \cite{lambrecht2009phd,meslin2010characterization} and SANS data \cite{bergner2010comparative} from the REVE campaign.}
  \label{monica_Z_vac_density_evolution.jp}
\end{figure}

\begin{figure}
 \centering
  \includegraphics[width=\columnwidth]{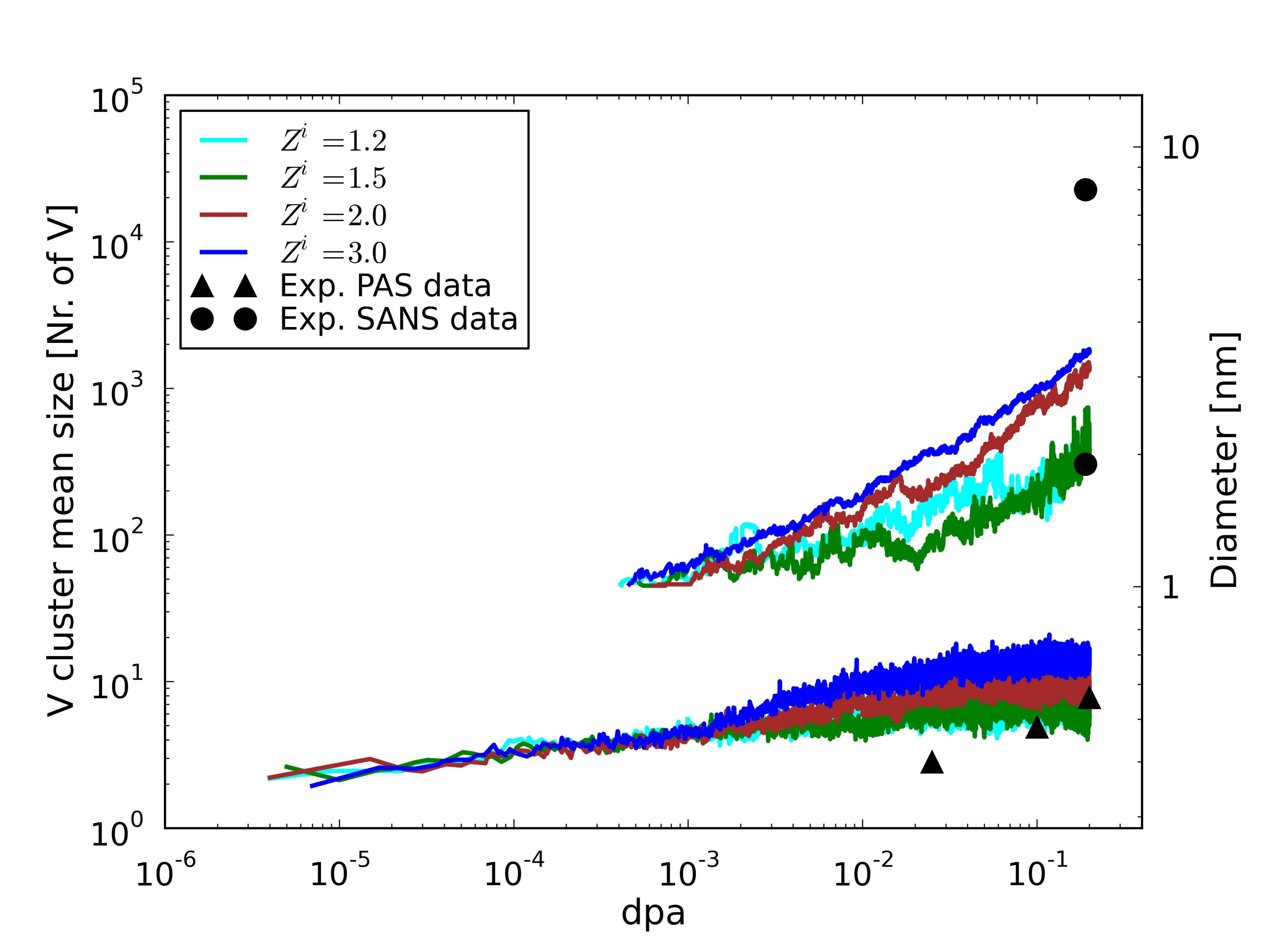}
    \caption{Vacancy cluster mean sizes versus dpa for different $Z^i$ values. Larger mean sizes are calculated using the SANS size resolution; smaller mean sizes are calculated using the PAS resolution. Results are compared with experimental PAS \cite{lambrecht2009phd,meslin2010characterization} and SANS data \cite{bergner2010comparative} from the REVE campaign. SANS points corresponds to the two peaks in the observed size distribution: the lower point is the major peak.}
  \label{monica_Z_vac_mean_size_evolution.pdf}
\end{figure}

The dislocation density plays a bigger role in this study than in \cite{jansson2013simulation,jansson2014okmc}, as the density here is larger, $\rho_d = (7\pm2)\cdot10^{13}$ m$^{-2}$. The effect of changing $Z^i$ is shown in Fig. \ref{monica_Z.pdf} and \ref{monica_Z_SIA_mean_size_evolution.pdf}for what concerns SIA clusters; the effect on the vacancy clusters is shown in Fig. \ref{monica_Z_vac_density_evolution.jp} and \ref{monica_Z_vac_mean_size_evolution.pdf}. Somehow, decreasing the value of $Z^i$ has similar effect to increasing the trapping energy for vacancies. A lower $Z^i$ increases the SIA visible cluster density lead to an earlier build up of SIA cluster density and also increases the growth of the SIA clusters. For vacancy clusters, the effect is reversed, so the density and cluster mean size are increased with increased $Z^i$, even though the effect is more significant for vacancy clusters visible by SANS than at the PAS size resolution. We get good agreement for the cluster density and the 
vacancy cluster mean size, using $Z^i=3.0$. For the SIA cluster mean size evolution, the same $Z^i$ underestimates the cluster mean sizes slightly, but remains acceptable. The mean size would be closer to experiment with smaller values of $Z^i$, but then the density of SIA clusters would be too large and the build up would start too early and with early saturation (Fig. \ref{monica_Z.pdf}).

\subsection{Impact of the concentration of carbon in the matrix}
\begin{figure}
 \centering
  \includegraphics[width=\columnwidth]{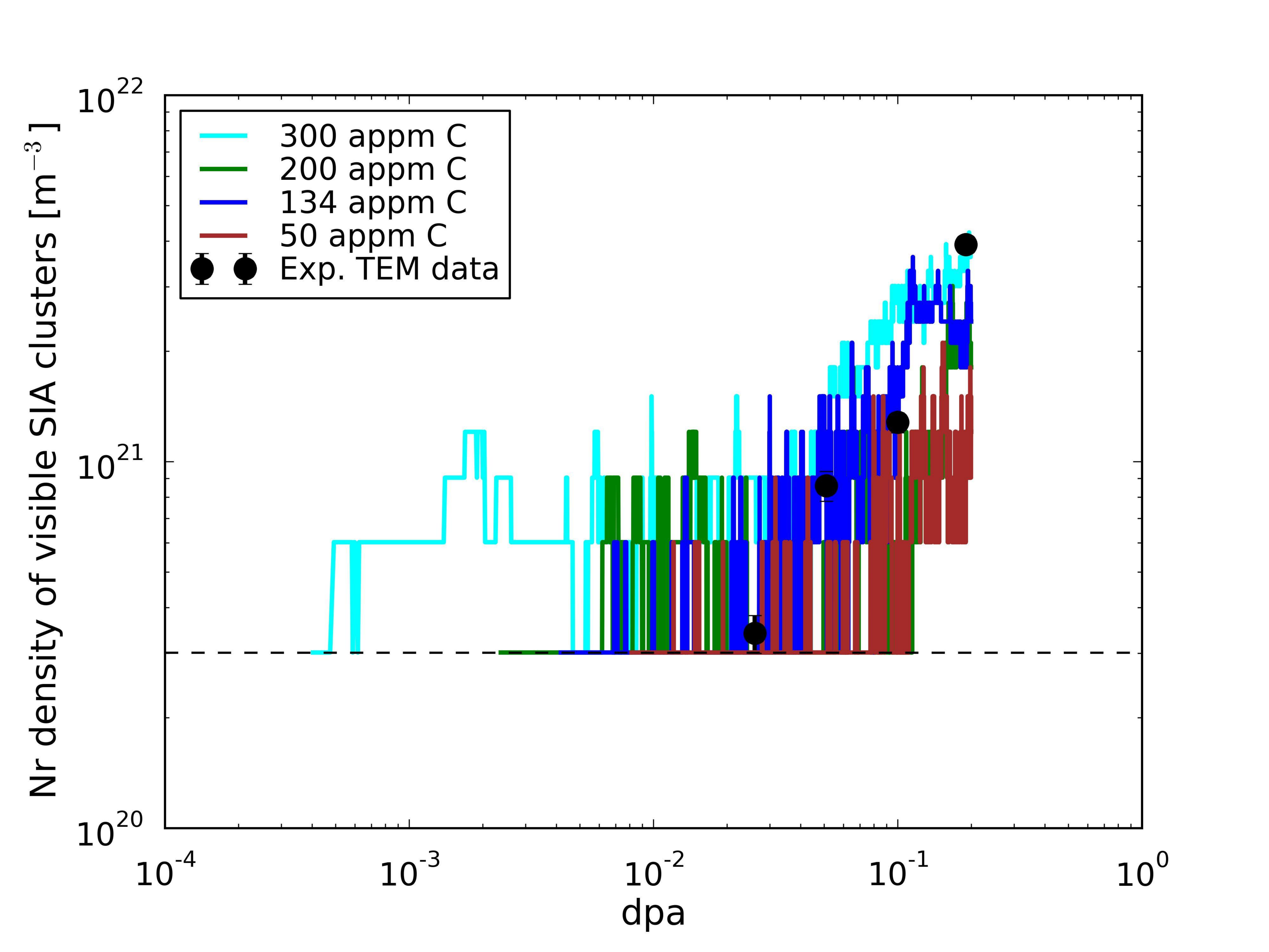}
    \caption{Visible SIA cluster density versus dpa for different C concentrations. The experimental TEM data are from \cite{hernandez2010transmission}. The dotted line corresponds to one cluster in the simulated system.}
  \label{monica_C.pdf}
\end{figure}

\begin{figure}
 \centering
  \includegraphics[width=\columnwidth]{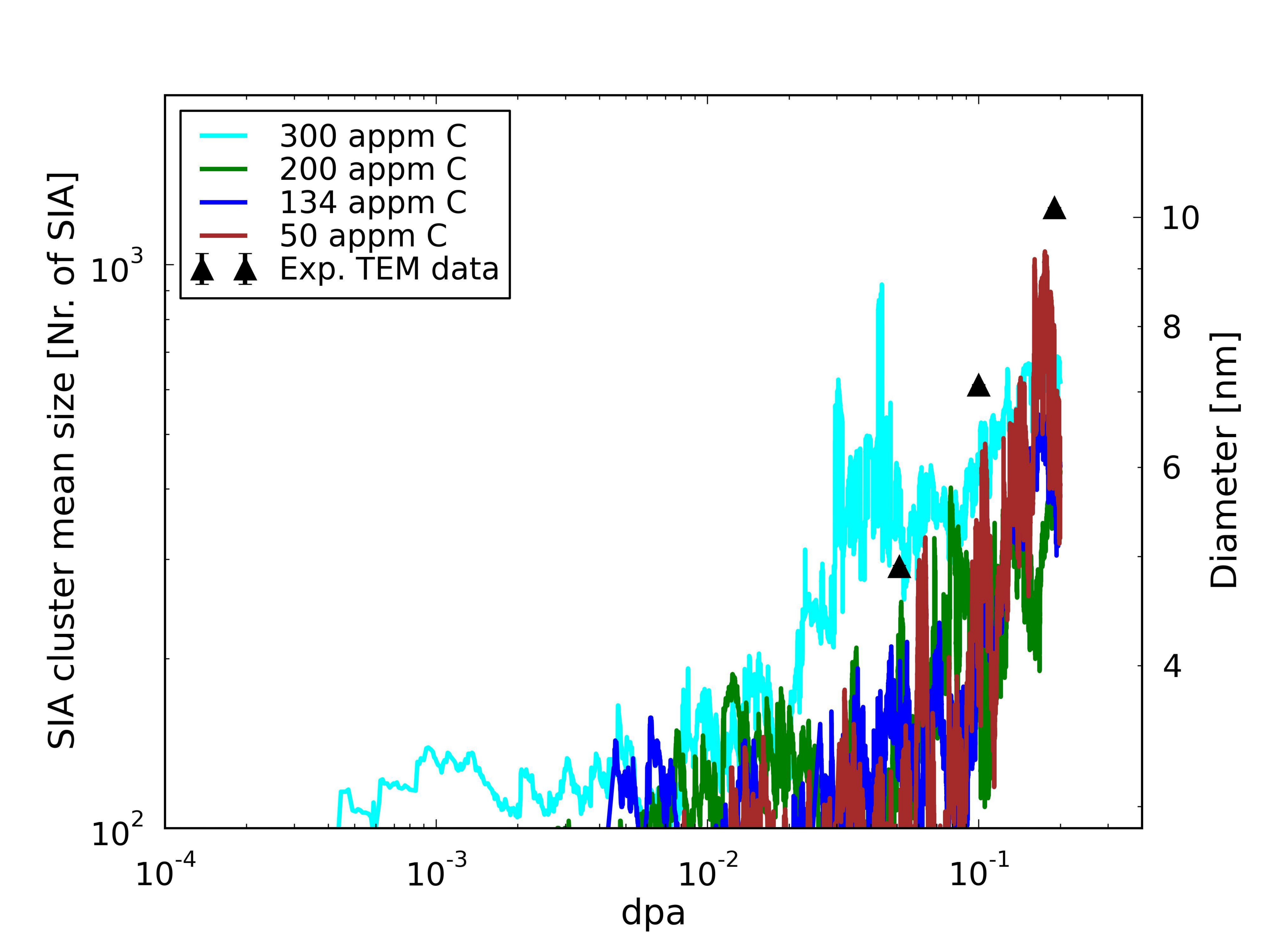}
    \caption{Visible SIA cluster mean size versus dpa for different C concentrations. The experimental TEM data are from \cite{hernandez2010transmission}.}
  \label{monica_C_SIA_mean_size_evolution.pdf}
\end{figure}

The impact of the C concentration was studied by trying different trap concentrations, as shown in Fig. \ref{monica_C.pdf}. Our reference experimental material is estimated to have a carbon concentration of 134 appm. Higher trap concentration gives a higher SIA cluster density, but even 200 appm of carbon still gives a fair agreement with the experimental data. Higher C concentrations also increase the mean SIA cluster sizes, as seen in Fig. \ref{monica_C_SIA_mean_size_evolution.pdf}.

\begin{figure}
 \centering
  \includegraphics[width=\columnwidth]{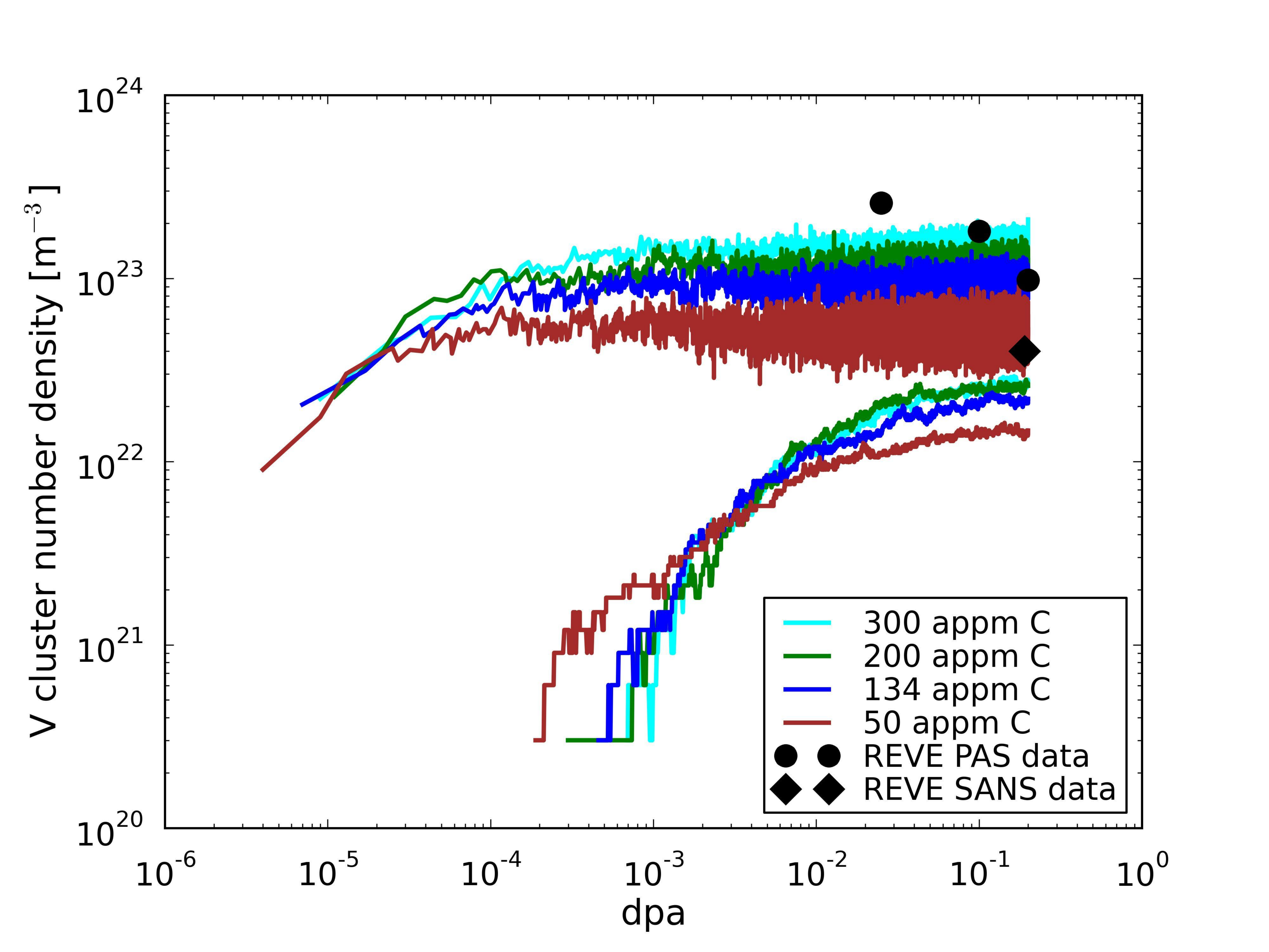}
    \caption{Vacancy cluster densities versus dpa for different C content. The higher densities are calculated with the PAS size resolution, whereas the lower densities with the same colour are calculated with the SANS size resolution. The data are compared with experimental PAS \cite{lambrecht2009phd,meslin2010characterization} and SANS data \cite{bergner2010comparative} from the REVE campaign.}
  \label{monica_C_vac_density_evolution.pdf}
\end{figure}

\begin{figure}
 \centering
  \includegraphics[width=\columnwidth]{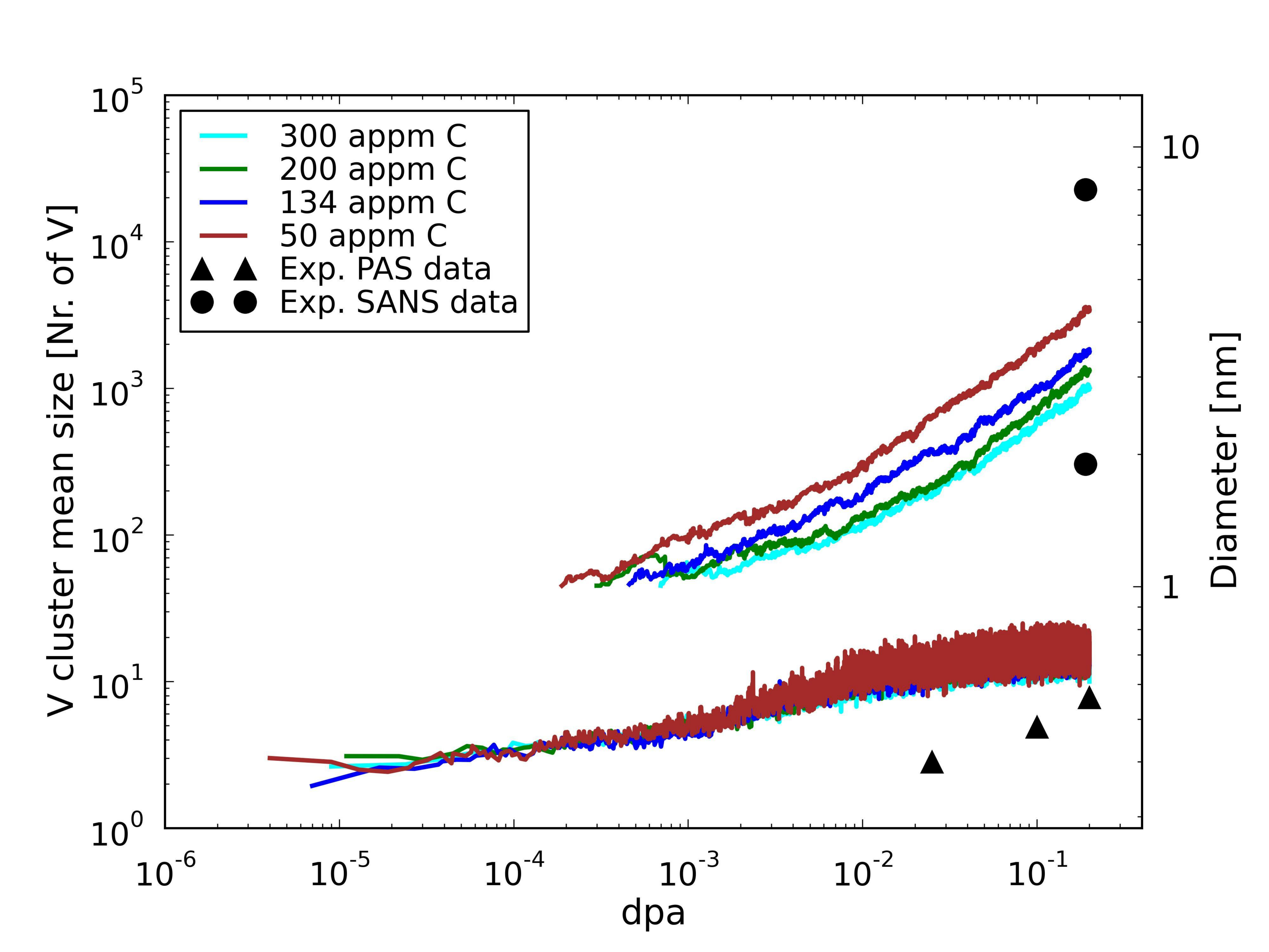}
    \caption{Vacancy cluster mean sizes versus dpa for different C content. Larger mean sizes are calculated using the SANS size resolution; smaller mean sizes are calculated using the PAS resolution. Results are compared with experimental PAS \cite{lambrecht2009phd,meslin2010characterization} and SANS data \cite{bergner2010comparative} from the REVE campaign. SANS points corresponds to the two peaks in the observed size distribution: the lower point is the major peak.}
  \label{monica_C_vac_mean_size_evolution.pdf}
\end{figure}

The vacancy density increases slightly with increased C concentration from $8.6\cdot10^{22}$ m$^{-3}$ (at 0.2 dpa) with 50 appm C to $2.09\cdot10^{23}$ m$^{-3}$ with 300 appm C (\textit{Cf.} Fig. \ref{monica_C_vac_density_evolution.pdf}). If only considering the cluster sizes observable by SANS, these densities increases as well slightly from $1.5\cdot10^{22}$ m$^{-3}$ (at 0.2 dpa) with 50 appm to $2.7\cdot10^{22}$ m$^{-3}$ with 300 appm C (\textit{Cf.} \ref{monica_C_vac_mean_size_evolution.pdf}). The experimental SANS value at 0.2 dpa with 134 appm C is $4.0\cdot10^{22}$ m$^{-3}$, so a higher density of C concentration slightly improves the agreement with the experimental SANS data. The density and mean size evolutions are all very similar, with early saturations for all four C concentrations.

The average vacancy cluster size for clusters visible by PAS are very similar for the four different C concentrations; between 0.61 and 0.67 nm and thus in good agreement with the experimental PAS value, 0.6 nm. No clear trend can be observed. For clusters visible by SANS, the mean sizes at 0.2 dpa goes from 4.23 nm with 50 appm C to 2.83 nm with 300 appm C. A larger C density appears to lead to more nucleation and less recombination of vacancy clusters, which leads to higher vacancy cluster densities, but slower growth and thus smaller mean size in the SANS size category, at the dose considered here. Less recombinations also promotes the growth of SIA clusters. The main peak in the SANS bimodal distribution is at a size of 1.9 nm. The agreement of the mean cluster sizes with SANS data thus improves with higher concentration of C, suggesting that the actual content of interstitial impurities in the material might have been higher than reported. It is also quite possible that we have underestimated the number 
of free C atoms, compared to the number of C$_2$V complexes. As C atoms are stronger traps than the C$_2$V complexes, less of the latter complexes gives more nucleation points.

\section{Discussion}\label{sec:discussion}

The evolution of the vacancy and SIA clusters under irradiation are highly dependent on each other, making it delicate to fine-tune parameters of physical significance but unknown value to satisfy the experimental data for both kinds of defects. Changing parameters for vacancies, such as the trapping energy for vacancy clusters, $E_t^v$, will have a significant effect on the SIA cluster evolution, as we have shown above. For the final set of parameters of our model, we had to fine-tune three parameters: the migration energy of invisible SIA clusters, $M^i$, ($N<N_{th}$), the energy to trap vacancy clusters at impurity atoms or other features, $E_t^v$ and the bias factor, $Z^i$. The chosen values had to make sense physically and make the simulations satisfy all the experimental data. In the fine-tuning process, it was seen that it is not straightforward to get at the same time good agreement for density and size. A compromise had to be found as, especially in the case of SIA clusters, values of the parameters 
that improve the mean size significantly increase the density, leading to too early build up and saturation. Although these differences are partially compensated by assuming that the content of interstitial species is higher than the nominal one, quite obviously the model has inherent limitations that we shall discuss in what follows.

% SIA

One strong limitation is given by the fact that we consider in the model, for the sake of simplicity, a single population of loops, with different features depending on size. These features correspond to either the properties of $\langle100\rangle$ loops, of to the properties of a "mixture" of $\langle100\rangle$ and $1/2\langle111\rangle$ loops. 
We do not know the proportions of $\langle100\rangle$ and $1/2\langle111\rangle$-SIA clusters, nor how the former clusters are created.
Are they the result of the interaction between $1/2\langle111\rangle$ loops created in cascades, as predicted by interatomic potentials \cite{stoller1996point,stoller1997primary,stoller2000statistical,
stoller2000evaluation,stoller2004secondary,stoller2000role,marian2002mechanism,terentyev2006study}? Or do they appear as the result of spontaneous transformation from $1/2\langle111\rangle$ loops \cite{arakawa2006changes}? Or is there an influence of temperature on their stability that makes them produced directly in cascades, this effect being unknown to the interatomic potentials \cite{dudarev2008effect}? Discussing the plausibility of these possibilities goes beyond the scope of the present work. Given this uncertainty, pragmatically we assumed that the type of loop is determined by size: visible loops are $\langle100\rangle$, invisible ones are of unknown nature. For these, we make use of an effective value of migration energy $M^i=0.2$ eV for $6\leq N^i \leq N_{th}$, where $N_{th}$ is the threshold for visibility. Clearly, a fully physical model should include a proper description of both classes of loops. However, given the uncertainties hinted at above, the development of such a model would imply 
exploring one by one all possible mechanisms, with the use, unavoidably, of other equally uncertain assumptions. Before such a model can be reliably developed more insight into the physics of SIA clusters is needed, otherwise the increased complexity would not be necessarily rewarded by an increased reliability.

Another strong approximation in our model concerns the fact that we introduce traps that are conceptually associated with the presence of C and C-vacancy complexes trapping SIA clusters, as well as vacancy clusters, but the behaviour of which is only remotely connected with C atoms and the complexes that it forms with defects. To start with, the traps are introduced since the very beginning, while in reality they will first need to form. Secondly, the traps are rigorously immobile and unchanged in nature throughout the simulation, while C atoms are highly mobile at the temperature considered here: the stability of C-vacancy complexes is such that they might be continuously forming and dissociating, and it cannot be excluded that some mechanism of migration of these complexes might exist. The dynamic formation and disappearance of complexes is likely to affect the kinetics of the nanostructural evolution under irradiation and we should probably suppose that the traps we use are somehow more efficient than the 
actual traps would be. However, again, the removal of this simplifying assumption and the introduction of a complete description of the physics of the traps implies the knowledge of a large number of parameters and mechanisms that we currently know very poorly. So, a more complete model should also eventually make use of assumptions, while being much more complex, thereby making it questionable whether the increased level of physical detail would be rewarded by higher reliability.
For the parameterization of the traps, we use \textit{ab initio} values for the interaction of single C atoms with SIA clusters, $E_{t1}^i$, for sizes up to $N^i=4$, and also with vacancy clusters, $E_t^v$, up to size 6 (\textit{Cf.} Table \ref{table:M_i}). Above these sizes  the trapping energies were assumed, guessed or fitted. For the SIA traps, two populations of traps were used, representing the two dominating complexes that are able to trap SIA clusters: C and C$_2$V. Below size $N_{th}$, we can assume that the more mobile $1/2\langle111\rangle$-SIA clusters, assumed to be present in the "mixture", are more affected by the traps than the slowly moving $\langle100\rangle$ clusters, so the trapping energies are chosen to represent the interaction between the former SIA clusters and the C ($E_{t1}^i$) and C$_2$V ($E_{t2}^i$) complexes, respectively. A C atom binds to a $1/2\langle111\rangle$-SIA cluster with $\sim$0.6 eV \cite{terentyev2011interaction}. This value is thus used for $E_{t1}^i$ for $5\leq 
N^i\leq N_{th}$. Above size $N_{th}$, the interaction is assumed to be exclusively with $\langle100\rangle$ and we used the MD result for the 
binding energy between said cluster and the C atom, $E_{t1}^i=1.1$ eV \cite{anento2013unpublished}.

The second kind of SIA trap is supposed to represent C$_2$V complexes. We do not know the exact values for the binding between these complexes and small SIA clusters. We used for the smallest sizes (1--4) $E_{t2}^i=0.6$ eV. For $N^i\geq 5$, but $<N_{th}$, we fitted the trapping energy to be $E_{t2}^i=1.2$ eV, which can be seen as an intermediate value between the binding energy between a C$_2$V cluster and a $1/2\langle111\rangle$-SIA cluster, 1.5 eV, or a $\langle100\rangle$-SIA cluster, 0.6 eV. We assume SIA clusters larger than $N_{th}$ to be of $\langle100\rangle$ type, and $E_{t2}^i = 0.6$ eV corresponds to the binding energy between C$_2$V complexes and these clusters \cite{anento2013unpublished}.

For the simulations at lower temperature, 343 K, reported in \cite{jansson2013simulation, jansson2014okmc}, no trapping of vacancies above size 6 was used. In this study at 563 K, we found that a trapping energy of $E_t^v=0.4$ eV is necessary. The higher mobility of vacancy clusters at the higher temperature makes the traps more important. At low temperature, the vacancy clusters are so slowly moving that traps have no effect. The vacancy traps represent C and C$_2V$ complexes, even though we assume them both to have the same trapping energies for vacancies. We do not have any knowledge of calculations of the binding energy between vacancy clusters above size 6 and C atoms, making fitting the only option.

% The Z
The bias factor for SIA sinks, $Z^i$, as defined in Eq. \eqref{eq:sink_radius}, proved to be an important parameter in our model. The bias factor takes into account the larger strain fields of SIA clusters, as compared to vacancy clusters. In rate theory calculations, the value chosen has usually been between 1.3 and 1.5. However, in order to lower the SIA density and still keep the general trend according to the experimental data, we found that a much higher value of $Z^i=3.0$ was needed. 

It is hard to find a physical explanation for why such a high value is needed and perhaps it is merely a result of other approximations of the model. Possibly, the assumption of immobile traps is responsible here.  Indeed, it is as if the C atoms, represented by traps, were immobile in our model. It is possible that several mobile C atoms could attach to the same vacancy or SIA cluster, thereby reducing the number of nucleation points for cluster growth. In our model, the number of traps is constant and given by the nominal C content in the material used in the experiment. We might thus be overestimating the number of nucleation points and, by increasing the sink strength for the SIA sinks by a large value of $Z^i$, we slow down the nucleation and clustering process at low dpa values. At higher dpa, the clusters are more dominating than the pre-existing sinks and new SIA defects are more likely to contribute to the cluster growth than to disappearing at the sinks. The main effect of a high $Z^i$ is thus a 
delay of the visible SIA cluster growth. It can be likewise used to explore the effect of different dose rates or high fluence. These studies will be the subject of forthcoming papers.

\section{Summary and conclusions}\label{sec:conclusions}

We have developed a physical model that describes nanostructural evolution under irradiation in Fe-C alloys versus dose, at temperatures in the range of the operation conditions of reactor pressure vessel steels. The model reproduces fairly well experimental data, using only a few calibration parameters of clear physical meaning. 
The main assumptions of the model, based on a combination of experimental and theoretical results, are: 
\begin{itemize}
 \item C atoms and complexes involving C atoms and vacancies, mainly C$_2$V, act as traps for SIA clusters and their effect can be described in first approximation as immobile traps to which a given trapping energy is associated that depends on the size of the trapped cluster. 
 
 \item While visible SIA loops have $\langle100\rangle$ Burgers vector, those invisible to the electronic microscope include loops of both $\langle100\rangle$ and $1/2\langle111\rangle$ type, and this fact can be taken into account by using effective migration parameters for invisible clusters. 
\end{itemize}

This model can be used as a starting point to add, explicitly or effectively, the effect of substitutional solute atoms found in reactor pressure vessel steels and known to be responsible for their hardening and embrittlement, such as Cu, Ni and Mn.

\section*{Acknowledgement}

This work was carried out as part of the PERFORM60 project of the 7th Euratom
Framework Programme, partially supported by the European Commission, Grant
agreement number FP7-232612.


%% References


\bibliographystyle{model1a-num-names}
%\bibliographystyle{apsrev} % The article titles are removed.
% \bibliography{vjansson,vjansson_publications}
\bibliography{/home/phys-data/people/vjansson/Beam/Articles/vjansson.bib,/home/phys-data/people/vjansson/Beam/Articles/vjansson_publications.bib}






% \end{thebibliography}

\end{document}